\documentclass[superscriptaddress,prd,showpacs,twocolumn,showpacs,nofootinbib]{revtex4}
\usepackage{float}
\usepackage{graphicx}
\usepackage{color}
\usepackage{epsf}
\usepackage{bm}
\usepackage{amsmath}
\usepackage{amsfonts}
\usepackage{amssymb}
\usepackage{epstopdf}%
\newcommand{\lp}{\left(}
\newcommand{\rp}{\right)}
\newcommand{\lb}{\left[}
\newcommand{\rb}{\right]}

\newcommand{\bea}{\begin{eqnarray}}
\newcommand{\eea}{\end{eqnarray}}
\newcommand{\be}{\begin{equation}}
\newcommand{\ee}{\end{equation}}

\newcommand{\mc}[1]{\mathcal{#1}}
\newcommand{\f}[2]{\frac{#1}{#2}}

\begin{document}

\title{Weyl type $f(Q,T)$ gravity, and its cosmological implications}
\author{Yixin Xu}
\email{xuyx27@mail2.sysu.edu.cn}
\affiliation{School of Physics, Sun Yat-Sen University, Xingang Road, Guangzhou 510275, People's
Republic of China}
\author{Tiberiu Harko}
\email{tiberiu.harko@aira.astro.ro}
\affiliation{Astronomical Observatory, 19 Ciresilor Street,  Cluj-Napoca 400487, Romania,}
\affiliation{Department of Physics, Babes-Bolyai University, Kogalniceanu Street,
Cluj-Napoca 400084, Romania}
\affiliation{School of Physics, Sun Yat-Sen University, Xingang Road, Guangzhou 510275, People's
Republic of China}
\author{Shahab Shahidi}
\email{s.shahidi@du.ac.ir}
\affiliation{School of Physics, Damghan University, Damghan,
	41167-36716, Iran}
\author{Shi-Dong Liang}
\email{stslsd@mail.sysu.edu.cn}
\affiliation{School of Physics, Sun Yat-Sen University, Xingang Road, Guangzhou 510275, People's
Republic of China}
\affiliation{State Key Laboratory of Optoelectronic Material and Technology, Guangdong Province Key Laboratory of Display Material and Technology, Guangzhou, People's Republic of China}

\begin{abstract}
We consider an $f(Q,T)$ type gravity model in which the scalar non-metricity $Q_{\alpha \mu \nu}$ of the space-time is expressed in its standard Weyl form, and it is fully determined by a vector field $w_{\mu}$. The field equations of the theory are obtained under the assumption of the vanishing of the total scalar curvature, a condition which is added into the gravitational action via a Lagrange multiplier.  The gravitational field equations are obtained from a variational principle,  and they explicitly depend on the scalar nonmetricity and on the Lagrange multiplier. The covariant divergence of the matter energy-momentum tensor is also determined, and it follows that the nonmetricity-matter coupling leads to the nonconservation of the energy and momentum. The energy and momentum balance equations are explicitly calculated, and the expressions of the energy source term and of the extra force are found. We investigate the cosmological implications of the theory, and we obtain the cosmological evolution equations for a flat, homogeneous and isotropic geometry, which generalize the Friedmann equations of standard general relativity. We consider several cosmological models by imposing some simple functional forms of the function $f(Q,T)$, and we compare the predictions of the theory with the standard $\Lambda$CDM model.
\end{abstract}

\maketitle

\tableofcontents

\section{Introduction}

The birth of general relativity as a result of the works by Einstein and Hilbert \cite{Ein1,Hilbert,Ein2} had a tremendous impact not only on physics and cosmology, but also on mathematics. In their works Einstein and Hilbert made an extensive use in their work of the Riemannian geometry \cite{Riemm}, in which a spacetime can be endowed with a metric and an affine structure, determined by a metric tensor $g_{\mu \nu}$ and a connection $\Gamma ^{\alpha}_{\mu \nu}$, respectively. The geometric and gravitational properties of the space time are described by the curvature tensor $R^{\mu }_{\nu \sigma \lambda }$ and its contraction, from which the Einstein tensor is constructed.

 Very soon after the emergence of general relativity, Weyl \cite{Weyl} did propose in 1918 an extension of Riemannian geometry , which he used for physical applications to develop the first unified theory of gravity and electromagnetism, in which the nonmetricity of the spacetime generated the electromagnetic field. Weyl's unified theory was severely criticized by Einstein, leading essentially to its abandonment for more than a half century. In the meantime another important development took place in differential geometry, and it was based on the introduction of the concept of torsion  \cite{Car1}. This led to an important extension of general relativity \cite{Car2,Car3,Car4}, which is called the Einstein-Cartan theory \cite{Hehl1}. From a physical point of view in the Einstein-Cartan theory the torsion field $T^{\mu }_{\sigma \lambda }\neq 0$ is identified with the spin density of the matter \cite{Hehl1}.

   A third independent mathematical and physical development of the gravitational field theories was initiated  by the work of Weitzenb\"{o}ck \cite{Weitz}, who introduced some geometrical structures  that are known presently as the Weitzenb\"{o}ck spaces. A Weitzenb\"{o}ck space is described by the properties $\nabla _{\mu }g_{\sigma \lambda }= 0$, $T^{\mu }_{\sigma \lambda }\neq 0$, and $R^{\mu }_{\nu \sigma \lambda }=0$, respectively. The  Weitzenb\"{o}ck space reduces  to a Euclidean manifold when $T^{\mu }_{\sigma \lambda }= 0$. On the other hand in a Weitzenb\"{o}ck manifold $T^{\mu }_{\sigma \lambda }$ takes different values  in different regions of the manifold. Since the Riemann curvature tensor identically vanishes in a  Weitzenb\"{o}ck manifold,  these  geometries have the key feature of distant parallelism, known also as teleparallelism or absolute parallelism. In physics Einstein was the first to apply Weitzenb\"{o}ck type space-times by proposing a unified teleparallel theory of gravitation and electromagnetism \cite{Ein}.

 The basic idea in the teleparallel  formulation of gravity is to substitute the metric $g_{\mu \nu}$ of the space-time manifold, representing the basic geometrical variable describing the gravitational field, by a set of tetrad vectors $e^i_{\mu }$. Then one can use the torsion tensor, generated by the tetrad fields,  to completely describe gravitational phenomena, with the curvature replaced by the torsion. Hence this approach leads to the so-called teleparallel equivalent of General Relativity (TEGR), which was proposed initially in \cite{TE1,TE2,TE3}, and presently it is also known as the $f(\mathbb{T})$ gravity theory, where $\mathbb{T}$ is the torsion scalar. The basic property of teleparallel, or $f(\mathbb{T})$ type theories, is that torsion exactly balances curvature, with the important result that the space-time turns into a flat manifold. Another important property of the $f(\mathbb{T})$ type gravity theories is that the gravitational field is described by  second order differential equations, a situation essentially different from other modified gravity theories, where, like, for example, in $f(R)$ gravity, the field equations in the metric approach are of fourth order \cite{bookHL}. A detailed analysis of teleparallel theories  is presented in \cite{book1}.  $f(\mathbb{T})$ gravity theories had been intensively used for the study of the cosmological evolution and of the astrophysical processes. They can provide a physical and geometrical explanation for the late-time accelerating expansion of the Universe, without the necessity of introducing a cosmological constant, or the dark energy \cite{TE4,TE5,TE6,TE7,TE8,TE9,TE10,TE11,TE12,TE13,TE14,TE15,TE16,TE17,TE18,TE19,TE20,TE21,TE22, TE23}.

 The Weyl geometry did not attract much attention in its first 50 years of existence. However, this situation changed after 1970, with the physicists beginning to gradually explore its interesting physical and mathematical consequences at both microscopic and macroscopic levels (for a very detailed description of the applications of Weyl geometry in physics see \cite{Scholz}.

 An interesting extension of Weyl gravity was proposed by Dirac \cite{Dirac1,Dirac2}. With the use of a real scalar field $\beta$ of weight $w(\beta)=-1$, and by constructing the electromagnetic field tensor $F_{\mu \nu}$ from the Weyl curvature, Dirac adopted as the gravitational Lagrangian the expression
 \be
 L=-\beta ^2R+kD^{\mu}\beta D_{\mu}\beta +c\beta ^4+\frac{1}{4}F_{\mu \nu}F^{\mu \nu},
 \ee
 where $k=6$ is a constant. This Lagrangian is conformally invariant. The cosmological implications of a slightly modified Dirac model were investigated in \cite{Rosen}. In \cite{Isrcosm} the evolution of a Universe described by the Weyl-Dirac type Lagrangian
 \bea
 L&=&W^{\lambda \rho}W_{\lambda \rho}-\beta ^2R+\sigma \beta ^2w^{\lambda}w_{\lambda}+2\sigma \beta w^{\lambda}\beta _{,\lambda}+\nonumber\\
&& (\sigma +6)\beta _{,\rho}\beta_{,\lambda }g^{\rho \lambda}+2\Lambda \beta ^4+L_m,
 \eea
was considered, where $W_{\mu \nu}$ is the Weyl length curvature tensor, constructed from the Weyl connection vector $w_{\mu}$, $\beta $ is the Dirac scalar field, while $\sigma$ and $\Lambda$ are constants. It turns out that in this model matter is created by Dirac’s gauge function at the beginning of the Universe, while in the dust dominated period Dirac’s gauge function gives rise to dark energy that causes the late time cosmic acceleration.

 Weyl's geometry can be extended naturally to include torsion. The corresponding geometry is called the Weyl-Cartan geometry, and it was extensively studied from both physical and mathematical points of view \cite{WC1,WC2,WC3,WC4,WC5,WC6,WC7,WC8, WC9}. For a review of the geometric properties and of the physical applications and of the Riemann-Cartan and Weyl-Cartan space-times  see \cite{Rev}.

 In the geometric and physical framework of the Weyl-Dirac theory torsion was included  in \cite{Isr1, Isr2,Isr3}, leading to  a Lagrangian
of the type
\begin {eqnarray}
L &=& W^{\mu\nu}W_{\mu\nu}-\beta^2 R+
\beta^2(k-6)w_\mu w^\mu+ \nonumber\\
&&2(k-6)\beta w^\mu
\beta_{,\mu}+k\beta_{,\mu} \beta_{,\underline\mu}
+8\beta\Gamma^\alpha_{\,\left[\lambda\alpha\right]}
\beta_{,\underline \lambda}
+\nonumber\\
&&\beta^2(2\Gamma^\alpha_{\,\left[\mu\lambda\right]}
\Gamma^\lambda_{\,\left[{\underline\mu}\alpha\right]}
-\Gamma^\alpha_{\,\left[\sigma\alpha\right]}
\Gamma^\omega_{\,\left[{\underline\sigma}\omega\right]}
+\Gamma^\alpha_{\,\left[\mu\lambda\right]}
\Gamma^\omega_{\,\left[\underline\mu\underline\lambda\right]}
g_{\alpha\omega}+\nonumber\\
&&8\Gamma^\alpha_{\,\left[\sigma\alpha\right]}w^\sigma)
 +4 W_{\mu\nu;\alpha}\Gamma^\alpha_{\,\left[
\underline\mu\underline\nu\right]} +2\Lambda
\beta^4+ L_{matter},
\end{eqnarray}
where the torsion tensor $\Gamma^\lambda_{\,\left[\,\mu\nu\right]} \rightarrow
\overline{\Gamma}^\lambda_{\,\left[\,\mu\nu\right]}=
\Gamma^\lambda_{\,\left[\,\mu\nu\right]}$ is gauge invariant, from which one can also construct a gauge covariant (in the sense of Weyl) general relativistic massive electrodynamics.

An extension of the teleparallel gravity models, called Weyl-Cartan-Weitzenb\"{o}ck gravity, was proposed in \cite{WCW}.
The action of this model can be formulated in terms of the dynamical variables $\left(g_{\mu \nu }, w_{\mu
},T^{\lambda}_{~\mu\nu}\right)$ as
\bea
S&=&\int
d^4x\sqrt{-g}\bigg(R+T^{\mu\alpha\nu}T_{\mu\alpha\nu}+2T^{\mu\alpha\nu}T_{
\nu\alpha\mu}-\nonumber\\
 &&4T_\mu T^\mu-\f{1}{4}W^{\mu \nu}W_{\mu \nu}+\beta \nabla_\mu T \nabla^\mu T-6w_\mu
w^\mu+\nonumber\\
&&8w_\mu T^\mu+L_m\bigg),
\eea
where $\beta $ is a constant, and $W_{\mu\nu}=\nabla_\nu w_{\mu}-\nabla_\mu w_{\nu}$, respectively.
In the Weyl-Cartan-Weitzenb\"{o}ck theory, the condition of the vanishing of the sum of the curvature and torsion scalar,
\be
R+T^{\mu\alpha\nu}T_{\mu\alpha\nu}+2T^{\mu\alpha\nu}T_{\nu\alpha\mu}-4T_\mu
T^\mu=0,
\ee
is imposed in a background  Weyl-Cartan type space-time, and it leads to a gravitational action of the form
\bea
S&=&\int d^4x \sqrt{-g}\bigg(-\f{1}{4}W^{\mu \nu}W_{\mu \nu}+\beta \nabla_\mu T
\nabla^\mu T-6w_\mu w^\mu\nonumber\\
&&+8w_\mu T^\mu+L_m\bigg).
\eea

An important difference with respect to the standard teleparallel theories is that the model is not formulated in a flat Euclidian geometry, but in a four-dimensional curved space-time. From the Weyl-Cartan-Weitzenb\"{o}ck theory a purely geometrical description of dark energy can be obtained, leading to a cosmological model in which the late time acceleration of the Universe is fully determined by the geometrical properties of the space-time.   The Weyl-Cartan-Weitzenb\"{o}ck  and the teleparallel gravity was extended in \cite{WCW1}, with the Weitzenb\"{o}ck condition in a Weyl-Cartan geometry  inserted into the gravitational action via a Lagrange multiplier. The action for this theory is
\begin{align}
S&=\f{1}{\kappa^2}\int d^4x\sqrt{-g}\bigg[-\f{\kappa^2}{4}W_{\mu\nu}W^{\mu\nu}-6w_\nu w^\nu+8 w_\nu T^\nu\nonumber\\
&+(1+\lambda)\big(R+T^{\mu\alpha\nu}T_{\mu\alpha\nu}+2T^{\mu\alpha\nu}T_{\nu\alpha\mu}-4T_\mu T^\mu\big)\nonumber\\
&+\beta\nabla_\mu T\nabla^\mu T+\alpha T_{\mu\nu}T^{\mu\nu}+\kappa^2L_m\bigg],
\end{align}
 where $\alpha$ and $\beta $ are constants, and $\lambda$ is the Lagrange multiplier. Both decelerating and accelerating cosmological models can be obtained from the theory.

 The above theoretical investigations suggests that general relativity can be represented in (at least) two mathematically equivalent geometric formalisms: the curvature formulation (in which the torsion and the nonmetricity identically vanish), and the teleparallel formulation, in which the curvature and the nonmetricity vanish identically, respectively.

  But a third equivalent geometric representation of general relativity is also possible. In this formulation the properties of the gravitational interaction are described geometrically by the nonmetricity $Q$ of the metric, which defines the variation of the length of a vector during the parallel transport around a closed loop. This approach is called the symmetric teleparallel gravity, and it was initially developed in \cite{Nester}. Generally, the connection describing the geometry can be decomposed into the Levi-Civita connection, and a deformation one form, $\Gamma ^{\alpha}_{\;\;\beta}=\Gamma ^{\{\}\alpha}_{\;\;\;\;\;\beta}-A^{\alpha}_{\;\;\beta}$, where $A_{\alpha \beta}=K_{\alpha \beta}-Q_{\alpha \beta}/2-Q_{\gamma [\alpha \beta]}\theta ^{\gamma}$, where $K_{\alpha \beta}$ is the contorsion, while $Q_{\alpha \beta}$ is the nonmetricity defined as $Q_{\alpha \beta}=-Dg_{\alpha \beta}$. By adopting a teleparallel frame in which $\Gamma $ vanishes, and by imposing the condition of the vanishing of the torsion, it turns out that $Q_{\mu \nu \lambda}=-g_{\mu \nu,\lambda}$, and the deformation tensor takes the form of the Christoffel symbol $\gamma ^{\alpha}_{\beta \gamma}$, $A^{\alpha}_{\;\;\beta \gamma}=\gamma ^{\alpha}_{\beta \gamma}$. The corresponding gravitational action takes the form $L_g=\sqrt{-g}g^{\mu \nu}\left(\gamma ^{\alpha}_{\beta \mu}\gamma ^{\beta}_{\nu \alpha}-\gamma ^{\alpha}_{\beta \alpha}\gamma ^{\beta}_{\mu \nu }\right)$, which is equivalent to the Hilbert-Einstein Lagrangian. Moreover, the associated energy-momentum density in symmetric teleparallel gravity is the Einstein pseudotensor, which in this geometric formulation becomes a true tensor. For a review of teleparallel gravity see \cite{revs}.

  The symmetric teleparallel gravity approach was further extended into the $f(Q)$ gravity theory (also called coincident general relativity) in \cite{Lav}. After introducing the quadratic nonmetricity scalar $Q=-Q_{\alpha \beta \mu}Q^{\alpha \beta \mu}/4+Q_{\alpha \beta \mu}Q^{\beta \mu \alpha}/2+Q_{\alpha}Q^{\alpha}/4-Q_{\alpha}\bar{Q}^{\alpha}/2$,  where $Q_{\mu}=Q^{\;\;\alpha}_{\mu \;\;\alpha}$, and $\tilde{Q}^{\mu}=Q_{\alpha}^{\;\;\mu \alpha}$, defining the nonmetricity conjugate $P^{\alpha}_{ \;\;\mu \nu}$ as $P^{\alpha}_{ \;\;\mu \nu}=c_1Q^{\alpha \;\mu \nu}+c_2Q^{\;\;\alpha}_{(\mu \;\nu)}+c^3Q^{\alpha}g_{\mu\nu}+c_4\delta^{\alpha}_{(\mu}\bar{Q}_{\nu)}+\left(c_5/2\right)\left(\tilde{Q}^{\alpha}g_{\mu \nu}+\delta^{\alpha}_{(\mu }Q_{\nu)}\right)$, and defining the general quadratic form $\mathrm{Q}$ as $\mathrm{Q}=Q_{\alpha}^{\;\mu \nu}P^{\alpha}_{\;\mu \nu}$, the gravitational action of the theory can be written down as \cite{Lav}
  \be
  S=\int{d^nx\left[-\frac{1}{2}\sqrt{-g}\mathrm{Q}+\lambda _{\alpha}^{\;\beta \mu \nu}R^{\alpha}_{\;\beta \mu \nu}+\lambda _{\alpha}^{\;\mu \nu}T^{\alpha}_{\;\mu \nu}\right]}.
  \ee

  Such gravitational theories based on nonmetricity may also be called nonmetric gravity. Different physical  and geometrical properties of symmetric teleparallel gravity have been investigated in the past in a number of studies, with the interest for this type of theoretical approach to gravity rapidly increasing recently \cite{s1,s2,s3,s4,s5,s6,s7,s8,s9,s10,s11, s12, s12a, s13,s14,s15,s16,s17,s18,s19,s20, s21}.

  In the so-called "newer general relativity" class theories the propagation velocity of the gravitational waves around Minkowski spacetime and their potential polarizations  were considered in \cite{s9}. For symmetric teleparallel spacetimes the exact propagator for the most general infinite-derivative, even-parity and generally covariant theories was obtained in \cite{s10}. For different extensions of symmetric teleparallel gravity the propagation of gravitational waves was studied in \cite{s13}, and it was found that the speed and the polarization of the gravitational waves are the same as in general relativity. In the framework of Symmetric Teleparallel Geometry an approach based on the Noether Symmetry was applied to classify all possible quadratic, first-order derivative terms of the non-metricity tensor  in \cite{s14}. The cosmological implications of the $f(Q)$ theory and its observational constraints were investigated in \cite{s15} and \cite{s16}, respectively. In this geometric theory the accelerating expansion of the Universe is an intrinsic property, and there is no need to introduce the dark energy. The evolution of the cosmological perturbations in $f(Q)$ gravity was analyzed in \cite{s20}.

In \cite{s12} an extension of symmetric teleparallel gravity was considered  by introducing a new class of theories where the nonmetricity $Q$ is nonminimally coupled to the matter Lagrangian. The action of the theory is given by
\be
S=\int{d^4x\sqrt{-g}\left[\frac{1}{2}f_1(Q)+f_2(Q)L_m\right]},
\ee
where $f_1$ and $f_2$ are arbitrary functions of $Q$, and $L_m$ is the matter Lagrangian. This nonminimal coupling between matter and geometry implies the nonconservation of the energy-momentum tensor, and to the generation of an extra force in the geodesic equation of motion. The cosmological solutions obtained for some specific functional forms of the functions $f_1(Q)$ and $f_2(Q)$ lead to  accelerating evolutions at late times.

 The most general extension of the symmetric teleparallel gravity, in which the gravitational Lagrangian $L$ is given by an arbitrary function $f$ of the non-metricity $Q$ and of the trace of the matter-energy-momentum tensor $T$, with action
 \be
 S=\int{\left[\frac{1}{16\pi}f(Q,T)+L_m\right]\sqrt{-g}d^4x},
 \ee
was studied in \cite{s21}. Cosmological models constructed by using some simple functional forms of the function $f( Q, T)$ were investigated in detail, and it was shown that for all considered cases the Universe experiences an accelerating expansion, ending with a de Sitter type evolution.
Geometry - matter couplings do appear in some semiclassical approaches to quantum gravity, where, for example, one can consider an action containing a geometry-quantum matter coupling of the form $\int{RF\left(\left<f(\phi)\right>\right)_{\Psi}\sqrt{-g}d^4x}$, where $\Psi$ is the wave function, $F$ and $f$ are arbitrary functions, and $\left(\left<f(\phi)\right>\right)_{\Psi}=\left<\Psi (t)\right|f[\phi (x)]\left|\Psi (t)\right>$ \cite{sg3a}. By assuming that the quantum metric can be decomposed into the sum of a classical and of a fluctuating part, of quantum origin, respectively, the resulting theories also lead at the classical level to modified gravity models with geometry-matter coupling \cite{re9,re11,re14},

It is the main goal of the present investigation to consider a particular implementation of the $f(Q,T)$ gravity theory, which is based on the nonminimal coupling between the nonmetricity $Q$ and the trace $T$ of the matter energy-momentum tensor. More exactly, we will go to the framework of the proper Weyl geometry, and adopt for the nonmetricity $Q$ the explicit expression that follows from the non-conservation of the divergence of the metric tensor in this geometry, $\nabla _{\lambda}g_{\mu \nu}=-w_{\lambda}g_{\mu \nu}$. This approach allows the representation of the nonmetricity in terms of a vector field $w_{\mu}$, and the metric tensor. With the help of the vector $w_{\mu}$ one can construct an electromagnetic type tensor associated to it. In Weyl geometry the nonmetricity is completely determined by the magnitude of $w_{\mu}$. In order to obtain a full dynamical description of the gravitational field we need to add to the gravitational action two terms related to the energy and the mass of the vector field. Moreover, in order to follow the essence of the teleparallel approach to gravity, we will also consider the flat geometry constraint, by requiring that the scalar curvature in the Weyl geometry vanishes. This constraint is added to the gravitational action via a Lagrange multiplier.

  Once the gravitational Lagrangian and the geometric action are constructed, we can obtain the gravitational field equations in the usual way. By varying the action with respect to the metric tensor we obtain the general field equations describing gravitational phenomena in the Weyl geometry in the presence of a massive vector field, coupled to the matter energy-momentum tensor, in a globally flat geometry. By varying the action with respect to the vector field we obtain the By considering the covariant derivative of the field equations we obtain the divergence of the  matter energy-momentum tensor does not vanish in the present approach to the gravitational interaction.  The cosmological implications of the $f(Q,T)$ theory  are investigated for three classes of specific models. The  obtained solutions describe both accelerating and decelerating evolutionary phases of the Universe, and they indicate that the Weyl type $f(Q,T)$ gravity can be considered as an alternative and useful approach for the description of the early and late phases of cosmological evolution.

  The present paper is organized as follows. The gravitational action and the field equations of the Weyl type $f(Q,T)$ theory are obtained in Section~\ref{sect1}. The energy and momentum balance equations are derived in Section~\ref{sectbal}. The cosmological evolution equations for a flat Universe geometry and their implications are considered in Section~\ref{sect2}. Specific cosmological models corresponding to different choices of the functional form of  $f(Q,T)$ are investigated in Section~\ref{sect3}. We discuss and conclude our results in Section~\ref{sect4}. The mathematical details of the derivation of the field equations are presented in Appendix~\ref{EOMderive}. The alternative representation of the field equations is described in Appendix~\ref{appB}. 

\section{Field Equations of the Weyl type $f(Q,T)$ theory}\label{sect1}

In the present Section we briefly review the basic concepts of the Weyl geometry, we introduce the variational principle of the Weyl type $f(Q,T)$ theory, and we write down the corresponding gravitational field equations. The divergence of the energy-momentum tensor is also calculated, and the energy and momentum balance equations of the theory are obtained.

\subsection{Weyl geometry in a nutshell}

In Riemannian geometry, if we parallelly transport a vector $v$ along an infinitesimal loop, the variation of its component is given by \cite{WCW}
\begin{equation}
    \delta v^\mu =v^{\kappa} R^{\mu}_{\ \kappa \sigma \nu} s^{\sigma \nu},
\end{equation}
where $s^{\sigma \nu}$ is the area encircled  by the loop.
Since $R_{\mu \nu \lambda \rho}$ is anti-symmetric with respect to the first two indices, the length of this vector is preserved, so that
\begin{equation}\label{length}
    \delta \lp g_{\mu \nu} v^{\mu} v^{\nu}\rp=2v^{\kappa}v^{\nu}R_{\nu \kappa \sigma \rho }s^{\sigma \rho}=0
\end{equation}

In order to describe the simultaneous change of direction and length, Weyl generalized the Riemannian geometry by introducing an intrinsic vector field $w^\mu$ and a semi-metric connection,
\begin{equation}
    \bar{\Gamma}^\lambda _{\ \mu \nu}\equiv \Gamma^\lambda _{\ \mu\nu}+g_{\mu\nu}w^\lambda-\delta^{\lambda}_{\mu}w_\nu-\delta^{\lambda}_\nu w_\mu,
\end{equation}
where $\Gamma^\lambda _{\ \mu\nu}$ is the Christoffel symbol constructed with respect to the metric $g_{\mu\nu}$. The curvature of this semi-metric connection has a symmetric part as well as an anti-symmetric part,
\begin{equation}
    \bar{R}_{\mu\nu\alpha\beta}=\bar{R}_{(\mu\nu)\alpha\beta}+\bar{R}_{[\mu\nu]\alpha\beta},
\end{equation}
where:
\begin{equation}
\begin{split}
    \bar{R}_{[\mu\nu]\alpha\beta}&=R_{\mu\nu\alpha\beta}+2\nabla_{\alpha}w_{[\mu}g_{\nu]\beta}+2\nabla_{\beta}w_{[\nu}g_{\mu]\alpha}\\&
    +2w_{\alpha}w_{[\mu}g_{\nu]\beta}+2w_{\beta}w_{
    [\nu}g_{\mu]\alpha}-2w^2g_{\alpha[\mu}g_{\nu ]\beta},
\end{split}
\end{equation}
and
\begin{equation}
    \bar{R}_{(\mu\nu)\alpha\beta}=g_{\mu\nu}W_{\alpha\beta},
\end{equation}
respectively, where
\begin{equation}
    W_{\mu\nu}=\nabla_{\nu}w_{\mu}-\nabla_{\mu}w_{\nu},
\end{equation}
is the field strength tensor of the vector field, while $R_{\mu\nu\alpha\beta}$ is the Riemann curvature tensor associated to the metric $g_{\mu \nu}$ \cite{WCW}.

From Eq.~(\ref{length}) we immediately see the geometric meaning of $W_{\mu\nu}$,
\begin{equation}
    \delta|v|=|v|W_{\sigma\rho}s^{\sigma \rho},
\end{equation}
where $|v|$ denotes the length of the vector. The first contraction of the Weyl curvature tensor is given by,
\begin{equation}
\begin{split}
    \bar{R}^\mu_{\ \nu}\equiv \bar{R}^{\alpha \mu}_{\ \ \alpha \nu}&=R^\mu_{\ \nu}+2w^\mu w_\nu +3\nabla^\nu w_\mu-\nabla_\mu w^\nu \\&+\delta^\mu_\nu\left(\nabla_\alpha w^\alpha-2w^2\right),
    \end{split}
\end{equation}
where $R^\mu_\nu$ is the Ricci tensor constructed from Riemann tensor and the Levi-Civita connection. The scalar curvature is
\begin{equation}
    \bar{R}\equiv \bar{R}^\alpha_{\ \alpha}=R+6\left(\nabla_\mu w^\mu-w^2\right).
\end{equation}

In Riemannian geometry, the Levi-Civita connection is compatible with the metric, i.e., $\nabla_\alpha g_{\mu\nu}=0$. This is not the case for the semi-metric connection in Weyl geometry, where we have \cite{WCW}
\begin{equation}\label{Qtensor}
    \begin{split}
        \bar{Q}_{\alpha\mu\nu} & \equiv \bar{\nabla}_\alpha g_{\mu\nu}=\partial_\alpha g_{\mu\nu}-\bar{\Gamma}^\rho_{\ \alpha \mu}g_{\rho\nu}-\bar{\Gamma}^\rho_{\ \alpha\nu}g_{\rho\mu}\\
      &=2w_\alpha g_{\mu\nu}.
    \end{split}
\end{equation}

The scalar non-metricity plays a central role in our theory, and it is given by
\begin{equation}
    Q\equiv -g^{\mu\nu}\left(L^\alpha _{\ \beta\nu}L^\beta_{\ \nu\alpha}-L^{\alpha}_{\ \beta\alpha}L^{\beta}_{\ \mu\nu}\right),
\end{equation}
where $L^\lambda_{\ \mu\nu}$ is defined as,
\begin{equation}
    L^{\lambda}_{\ \mu\nu}=-\frac{1}{2}g^{\lambda\gamma}\left(Q_{\mu\gamma\nu}+Q_{\nu\gamma\mu}-Q_{\gamma\mu\nu}\right).
\end{equation}
Plugging Eq~(\ref{Qtensor}) into the expression above, we obtain the important relation
\begin{equation}\label{Q}
    Q=-6w^2.
\end{equation}

\subsection{The variational principle and the field equations}

With all the geometric preliminaries in place, we can move on to discuss the field theory itself.
We consider the following action
\begin{equation}\label{act}
\begin{split}
    S&=\int d^4x \sqrt{-g}\Big[	\kappa^2f(Q,T)-\frac{1}{4}W_{\mu\nu}W^{\mu\nu}\\
    &-\frac{1}{2}m^2 w_\mu w^\mu+\mathcal{L}_m \Big].
    \end{split}
\end{equation}

In Eq.~(\ref{act}) $\kappa^2\equiv1/16\pi G$, $m$ is the mass of the particle associated to the vector field, while $\mathcal{L}_m$ is the matter action.   The second and third terms in the action are the ordinary kinetic term and mass term of the vector field, respectively.
The dynamics of the gravitational field is characterized by this action together with a flat geometry constraint, through which we impose the vanishing of the total curvature of the Weyl space,
\begin{equation}\label{cons1}
    \bar{R}=0.
\end{equation}

We impose this constraint by adding a Lagrange multiplier in the gravitational action, which becomes
\begin{equation}
\begin{split}
    S&=\int d^4x \sqrt{-g}\Bigg[\kappa^2f(Q,T)-	\frac{1}{4}W_{\mu\nu}W^{\mu\nu}\\
    &-\frac{1}{2}m^2 w_\mu w^\mu+\lambda(R+6\nabla_\alpha w^\alpha-6w_\alpha w^\alpha) + \mathcal{L}_m \Bigg].
    \end{split}
\end{equation}

Varying the action with respect to the vector field, we obtain the generalized Proca equation describing the field evolution,
\begin{equation}\label{EOM1}
    \nabla^\nu W_{\mu \nu }-(m^2+12\kappa^2 f_Q+12\lambda)w_\mu=6\nabla_\mu \lambda.
\end{equation}

Comparing this equation with the standard Proca equation, we see that the effective dynamical mass of the vector field is given by
\begin{equation}\label{discussion1}
m^2_{\rm{eff}}=m^2+12\kappa^2f_Q+12\lambda.
\end{equation}
We can also see that the Lagrange multiplier field generates an effective current for the vector field. From quantum field theory, we know that the mass detected in experiments may deviate from the bare mass due to the existence of interaction. Eq.~(\ref{discussion1}) shows that in the Weyl type $f(Q,T)$ gravity, this deviation can also originate from the nontrivial structure of the spacetime.

Variation with respect to the metric field gives the following field equation (see Appendix~\ref{EOMderive} for the calculation details),
\bea\label{EOM2}
 \hspace{-0.9cm}    &&\f12\left( T_{\mu\nu}+S_{\mu\nu}\right)-\kappa^2f_T \left(T_{\mu\nu}+\Theta_{\mu\nu}\right)=-\frac{\kappa^2}{2}g_{\mu\nu}f\nonumber\\
 \hspace{-0.9cm}      &&-6\kappa^2f_Q w_\mu w_\nu+\lambda \left(R_{\mu\nu}-6w_\mu w_\nu+3g_{\mu\nu}\nabla_\rho w^\rho\right)\nonumber\\
 \hspace{-0.9cm}     && +3g_{\mu\nu} w^\rho \nabla_\rho \lambda   -6w_{(\mu}\nabla_{\nu)}\lambda
      +g_{\mu\nu}\Box\lambda -\nabla_\mu \nabla_\nu \lambda,
\eea
where we have defined,
\begin{align}
T_{\mu\nu}&\equiv-\frac{2}{\sqrt{-g}}\frac{\delta(\sqrt{-g}\mathcal{L}_m)}{\delta g^{\mu\nu}},
\end{align}
and
\begin{align}
f_T&\equiv \frac{\partial f(Q,T)}{\partial T},\quad f_Q\equiv \frac{\partial f(Q,T)}{\partial Q},
\end{align}
respectively. Also, we have introduced the quantity $\Theta _{\mu \nu}$, defined as
\begin{align}
\Theta_{\mu\nu}\equiv g^{\alpha\beta}\f{\delta T_{\alpha\beta}}{\delta g_{\mu\nu}}=g_{\mu\nu}\mathcal{L}_m-2T_{\mu\nu}-2g^{\alpha\beta}\f{\delta^2\mc{L}_m}{\delta g^{\mu\nu}\delta g^{\alpha\beta}}.
\end{align}

\emph{\emph{}}In the field equation above, $S_{\mu\nu}$ is the rescaled energy momentum tensor of the free Proca field,
\begin{equation}
\begin{split}
    S_{\mu\nu}=&- \frac{1}{4}g_{\mu\nu} W_{\rho \sigma} W^{\rho \sigma}+W_{\mu\rho}W_{\nu}^{\ \rho}\\
    &-\frac{1}{2}m^2 g_{\mu\nu}w_{\rho}w^{\rho}+m^2w_\mu w_\nu.
    \end{split}
\end{equation}

In terms of the Einstein tensor $G_{\mu \nu}=R_{|mu \nu}-g_{\mu\nu}R/2$ the field equations become
\begin{eqnarray}
\hspace{-0.4cm}&&R_{\mu \nu }-\frac{1}{2}Rg_{\mu \nu } =\frac{1}{2\lambda }%
\left( T_{\mu \nu }+S_{\mu \nu }\right) -\frac{\kappa ^{2}}{\lambda }%
f_{T}\left( T_{\mu \nu }+\Theta _{\mu \nu }\right) \nonumber\\
\hspace{-0.4cm}&&+\frac{\kappa ^{2}}{%
2\lambda }g_{\mu \nu }f
+6\frac{\kappa ^{2}}{\lambda }f_{Q}w_{\mu }w_{\nu }-\frac{3}{\lambda }%
g_{\mu \nu }w^{\rho }\nabla _{\rho }\lambda +\frac{6}{\lambda }w_{(\mu
}\nabla _{\nu )}\lambda \nonumber\\
\hspace{-0.4cm}&&-\frac{1}{\lambda }\left( g_{\mu \nu }\Box \lambda
-\nabla _{\mu }\nabla _{\nu }\lambda \right)
+6w_{\mu }w_{\nu }-3g_{\mu \nu }w^{2}.
\end{eqnarray}

Alternatively, the gravitational field equations can be reformulated in the form
\begin{align}\label{altf}
\f12&\left(T_{\mu\nu}+S_{\mu\nu}\right)-\kappa^2f_T\left(T_{\mu\nu}+\Theta_{\mu\nu}\right)=-\frac{\kappa^2}{2}g_{\mu\nu}f\nonumber\\&-6\kappa^2f_Q w_\mu w_\nu+  g_{\mu\nu}\nabla^\rho D_\rho \lambda-\nabla_\nu D_\mu\lambda\nonumber\\
&
+ g_{\mu\nu}w_\rho D^\rho \lambda-w_\mu D_\nu \lambda-3w_\nu D_\mu \lambda\nonumber\\
&
+\lambda \lp R_{\mu\nu}+2w_\mu w_\nu-2g_{\mu\nu}w^2+g_{\mu\nu}\nabla_\rho w^\rho+2\nabla_\nu w_\mu\rp,
\end{align}
where we have denoted $D_\mu=\nabla_\mu+2w_\mu$ (for the derivation of Eqs.~(\ref{altf}) see Appendix~\ref{appB}).

Taking the trace of both sides of Eq.~(\ref{altf}), we obtain first
\begin{align}
\f12\left(T+S\right)&-\kappa^2f_T\left(T+\Theta\right)=-2\kappa^2f-6\kappa ^2f_Q w^2 \nonumber\\&+3 \nabla^\rho D_\rho \lambda
+\lambda \lb R+6\lp \nabla_\rho w^\rho -w^2 \rp \rb.
\end{align}

Due to the flat geometry constraint, the term proportional to $\lambda$ in the above equation vanishes.
From Eq.~(\ref{EOM1}), we can derive the explicit form of the term $\nabla^\rho D_\rho \lambda$, and thus we obtain
\begin{align}
 \f12\left(T+S\right)&-\kappa^2 f_T\left(T+\Theta\right)=-2\kappa^2 f-6\kappa^2f_Q w^2\nonumber\\
&-\frac{m^2}{2}\nabla^\rho w_\rho -6\nabla^\rho(f_Q w_\rho).
\end{align}

From the expression of $S_{\mu\nu}$ we obtain $S=-m^2w^2$. Hence we have
\begin{equation}\label{simpEOM2}
    \begin{split}
    &\f12(1-2\kappa^2 f_T)T-\kappa^2f_T \Theta\\&
    =-2\kappa^2f +\kappa^2 Q f_Q-6\nabla^\rho(f_Q w_\rho)+\frac{m^2}{2}(w^2-\nabla_\rho w^\rho)\\
    &=-2\kappa^2f +\kappa^2 Q f_Q-6\nabla^\rho(f_Q w_\rho)+\frac{m^2R}{12},
    \end{split}
\end{equation}
where to obtain the last line we have used again the flat constraint $\bar{R}=0$.

\section{The energy and momentum balance equations}\label{sectbal}

After taking the covariant divergence of the metric field equation \eqref{EOM2}, and using equations \eqref{EOM1} and \eqref{discussion1}, one can obtain the conservation equation of the energy-momentum tensor as
\begin{align}\label{div0}
\nabla^\mu &T_{\mu\nu}=\frac{1}{1+2\kappa^2 f_T}\Big [ 2\kappa^2\nabla_\nu(pf_T)-\kappa^2f_T\nabla_\nu T\nonumber\\&-2\kappa^2T_{\mu\nu}\nabla^\mu f_T-m_{eff}^2(w^\mu W_{\nu\mu}-w_\nu\nabla_\mu w^\mu)\nonumber\\&-6w_\nu\Box\lambda-6W_{\nu\mu}\nabla^\mu\lambda+W_{\nu\mu}\nabla_\alpha W^{\mu\alpha}\nonumber\\&-12w_\nu w^\mu\nabla_\mu(\lambda+\kappa^2f_Q)\Big ].
\end{align}
By using the vector field Eq.~\eqref{EOM1}, we obtain
\begin{align}
 \nabla^\mu T_{\mu\nu}&=\frac{1}{1+2\kappa^2 f_T}\Big [2(\nabla_\nu\Box-\Box\nabla_\nu+G_{\mu\nu}\nabla^\mu)\lambda \nonumber\\&+6\nabla_\nu\lambda(w^2-\nabla_\mu w^\mu)+w_\nu(\nabla_\alpha\nabla_\mu\nabla^\alpha-\nabla_\mu\Box)w^\mu\nonumber\\&+\nabla^\alpha w^\mu(\nabla_\alpha W_{\nu\mu}+\nabla_\mu W_{\nu\alpha}+\nabla_\nu W_{\mu\alpha})\nonumber\\&+2\kappa^2\nabla_\nu(pf_T)-f_T\nabla_\nu T-2T_{\mu\nu}\nabla^\mu f_T\Big ].
\end{align}
Now, by simplifying the covariant derivatives and using the constraint equation Eq.~\eqref{cons1}, we find
\begin{align}\label{div}
    \nabla^\mu T_{\mu\nu}=\frac{\kappa^2}{1+2\kappa^2 f_T}\Big [ 2\nabla_\nu(pf_T)-f_T\nabla_\nu T-2T_{\mu\nu}\nabla^\mu f_T\Big ].
\end{align}
It should be noted that the energy-momentum tensor becomes conserved in the case $f_T=0$.

We consider the matter content of the gravitating system as represented by perfect fluid, and we take the energy momentum tensor as
\begin{equation}\label{perfT}
T_{\mu\nu}=(\rho+p)u_\mu u_\nu +p g_{\mu\nu},
\end{equation}
where $\rho$ is the total matter energy, and $p$ is the thermodynamic pressure, respectively. The four-velocity $u^\mu$ is the tangent vector of a particle's worldline, parameterized  by the arc length, and  hence satisfies the normalization condition $u_\mu u^\mu=-1$.
Taking the covariant derivative of Eq.~(\ref{perfT}), we obtain first
\begin{equation}\label{perfdivT}
\begin{split}
    &\nabla^\mu T_{\mu\nu}=\lp\nabla^\mu p+\nabla^\mu \rho \rp u_\mu u_\nu\\&+(p+\rho)\lp u_\nu\nabla^\mu u_\mu +u_\mu \nabla^\mu u_\nu \rp +\nabla_\nu p.
\end{split}
\end{equation}
Multiplying with $u^\nu$ both sides of the above relation, we obtain the energy balance equation,
\begin{equation}
    \begin{split}
        u^\nu \nabla^\mu T_{\mu\nu}=-u_\mu \nabla^\mu \rho-(p+\rho)\nabla^\mu u_\mu \equiv -\dot{\rho}-3H(p+\rho)
    \end{split}
\end{equation}
where we have used the relation $u^{\mu} \nabla_{\nu} u_{\mu}=0$, and we have introduced the Hubble function $H$, defined as $3H\equiv \nabla^\mu u_\mu$. The dot is defined as $u_\mu \nabla^\mu=d/ds$, with $s$ being the arc length along the worldline of the particle.

The energy source $\mathcal{S}$ in the gravitating system is given by
\begin{equation}
    \mathcal{S}\equiv \dot{\rho}+3H(p+\rho)=-u^\nu \nabla^\mu T_{\mu\nu}.
\end{equation}

Multiplying with the projection operator $h^{\nu\rho}\equiv g^{\nu\rho}+u^\nu u^\rho$ both sides of Eq.~(\ref{perfdivT}), we obtain the momentum balance equation as given by
\begin{equation}
    u^\mu \nabla_\mu u^\rho=\frac{d^2x^{\rho}}{ds^2}+\Gamma^{\rho}_{\mu\lambda}\frac{dx^\mu}{ds}\frac{dx^\lambda}{ds} =\frac{h^{\nu\rho}}{p+\rho}\lp\nabla^\mu T_{\mu\nu}-\nabla_\nu p \rp.
\end{equation}
From the equation above we can see that the quantity
\be
\frac{h^{\nu\rho}}{p+\rho}\lp\nabla^\mu T_{\mu\nu}-\nabla_\nu p \rp,
 \ee
 measures the deviation of a particle's worldline from a geodesic, and hence it should be interpreted as a generalized force,
\begin{equation}
    \mathcal{F}^\rho=-\frac{h^{\nu\rho}\nabla_\nu p}{p+\rho}+\frac{h^{\nu\rho}\nabla^\mu T_{\mu\nu}}{p+\rho}.
\end{equation}

Now, using Eq.~(\ref{div}), one can obtain the generalized force as
\begin{align}\label{force}
    \mathcal{F}^\rho=\frac{\kappa^2h^{\nu\rho}}{(\rho+p)(1+2\kappa^2 f_T)}\Big [&-\f{1}{\kappa^2}\nabla_\nu p+ 2p\nabla_\nu (pf_T)\nonumber\\&-f_T\nabla_\nu T-2T_{\mu\nu}\nabla^\mu f_T\Big ].
\end{align}

 Finally, we will also discuss briefly the divergence of energy-momentum tensor $S_{\mu\nu}$ of the Weyl vector field. By making use of the generalized Proca equation and of the Jacobi identity $\nabla_\nu W_{\rho\sigma}+\nabla_{\sigma}W_{\nu\rho}+\nabla_{\rho}W_{\sigma\nu}=0$, we obtain immediately
\begin{equation}\label{discussion2}
    \kappa^2 \nabla_\mu S^\mu _\nu =6\left[W_{\mu\nu}\nabla ^\mu \lambda -\left(\frac{m}{m_{\rm{eff}}}\right)^2w_{\nu}\Box \lambda \right].
\end{equation}

It follows that when the Lagrange multiplier $\lambda$ is a constant, the divergence of the Weyl vector field is zero.
 By using Eq.~(\ref{discussion1}) and (\ref{discussion2}), we can roughly classify the relation between space-time geometry and the vector particle into three categories:

\paragraph{Decoupled phase}

When $f$ has a trivial dependence on $Q$, and $\lambda$ is zero, the mass of the vector particle is equal to its bare mass, and the total number of particles is conserved. In this case, however, the evolution of space-time structure as well as the dynamic of the Weyl particles  are still mutually dependent.  Space-time and the vector particle are not entirely detached, and there is a mutual influence between them.
\paragraph{Weakly entangled phase} When $f$ has a trivial dependence on $Q$, and $\lambda $  is a nonzero constant, the mass of the particle is shifted but the total  particle number is still conserved. The weakly entangled phase is basically the same as the decoupled phase, since a constant shift in the mass cannot be observed experimentally.
\paragraph{Strongly entangled phase} When $f$ has a nontrivial dependence on $Q$, or $\lambda$ has nontrivial dependence on the space-time, the effective mass of the particle will not only be different from its bare mass, but may also change with time and position. More importantly, the vector particles can be created or annihilated from the space-time continuum.

\section{Cosmological evolution of the flat Friedmann-Robertson-Walker Universe in the Weyl type $f(Q,T)$ gravity}\label{sect2}

In the following we will proceed to the investigation of the cosmological applications of the Weyl type $f(Q,T)$ theory. We assume that the geometry of the Universe is described by the isotropic, homogeneous and spatially flat Friedmann-Robertson-Walker metric, given by
\begin{equation}
ds^2=-dt^2+a^2(t)\delta_{ij}dx^idx^j,
\end{equation}
where $a$ is the scale factor. Due to spatial symmetry, the vector field is taken to be of the form
\begin{equation}
    w_{\mu}=[\psi(t),0,0,0].
\end{equation}
Therefore $w^2=w_{\mu}w^{\mu}=-\psi ^2(t)$, giving $Q=-6w^2=6\psi ^2(t)$.

Moreover, we adopt a comoving coordinate system with $u^{\mu}=(-1,0,0,0)$. In this case, $u^\mu \nabla_\mu=d/dt$ and $H=\dot{a}/a$.
We also fix the Lagrangian of the perfect fluid to be $\mc{L}_m=p$.
As a result, we obtain   $$T^{\mu}_{\nu}=\text{diag}(-\rho,p,p,p),$$
and
\begin{equation}
\begin{split}
\Theta^\mu_{\nu}=\delta^{\mu}_{\nu}p-2T^{\mu}_{\nu}&=\text{diag}(2\rho+p,-p,-p,-p).
    \end{split}
\end{equation}

\subsection{The generalized Friedmann equations}

For the cosmological case the flat space constraint, and the generalized Proca equation can be represented as
\begin{align}\label{constraint}
    \dot{\psi}&=\dot{H}+2H^2+\psi^2-3H\psi,\\ \label{proca}
    \dot{\lambda}&=\left(-\frac{1}{6}m^2-2\kappa^2f_Q-2\lambda\right)\psi=-\frac{1}{6}m_{{\rm eff}}^2\psi,\\ \label{cons}
    \partial_i \lambda&=0.
\end{align}

From Eqs.~(\ref{EOM2}) we obtain the generalized Friedmann equations as
\begin{align}
      \kappa^2  f_T(\rho+p)&+\f12 \rho =\frac{\kappa^2}{2}f-\left(6\kappa^2f_Q+\frac{1}{4}m^2\right)\psi^2\nonumber\\
        &-3\lambda(\psi^2-H^2)-3\dot{\lambda}(\psi-H),
\end{align}
\begin{align}
       - \f12 p&=\frac{\kappa^2}{2}f+\frac{m^2\psi^2}{4}+\lambda(3\psi^2+3H^2+2\dot{H})\nonumber\\&+(3\psi+2H) \dot{\lambda}+\ddot{\lambda}.
\end{align}

 With the use of Eqs.~(\ref{constraint}) and Eq.~(\ref{proca}) we eliminate all the derivatives of  $\lambda$, and then we take the sum of the two equations above. Hence we obtain a simpler set of the cosmological evolution equations, given by
\begin{align}\label{simpfe1}
   \f12 \left(1+2\kappa^2f_T\right)\rho&+\kappa^2f_T p=\frac{\kappa^2}{2}f+\frac{m^2\psi^2}{4}\nonumber\\&+3\lambda\left(H^2+\psi^2\right)-\f12m_{eff}^2H\psi,
\end{align}
\begin{align}\label{simpfe2}
  \f12  \left(1+2\kappa^2f_T\right)\left(\rho+p\right)&=\frac{m_{eff}^2}{6}\left(\dot{\psi}+\psi^2-H\psi\right)\nonumber\\&+2\kappa^2\dot{f_Q}\psi-2\lambda\dot{H}.
\end{align}
By substituting $\dot{\psi}$ as given by Eq.~(\ref{proca}) in Eq.~(\ref{simpfe2}) we obtain
\bea
&&\f12  \left(1+2\kappa^2f_T\right)\left(\rho+p\right)=-2\lambda \left(1-\frac{m_{eff}^2}{12\lambda}\right)\dot{H}+\nonumber\\
&&\frac{m_{eff}^2}{3}\left(H^2+\psi ^2-2H\psi\right)+2\kappa ^2\dot{f}_Q\psi.
\eea

The energy balance equation can be obtained as
\begin{align}
\dot{\rho}+3H(\rho+p)&=\frac{1}{1+2\kappa^2f_T}\Big[ 2\kappa^2(\rho+p)\dot{f}_T-f_T(\dot{\rho}-\dot{p})\Big].
\end{align}

The generalized Friedmann equations (\ref{simpfe1}) and (\ref{simpfe2}) can be reformulated in an effective form as
\be\label{gen1}
3H^2=\frac{1 }{2\lambda}\left(\rho+\rho _{eff}\right),
\ee
\be\label{gen2}
2\dot{H}=-\frac{1}{2\lambda }\left(\rho +\rho _{eff}+p+p_{eff}\right),
\ee
where
\be\label{rhoeff}
\rho _{eff}=m_{{\rm eff}}^2H\psi+2\kappa ^2f_T\left(\rho +p\right)-\kappa ^2f-\frac{m^2\psi ^2}{2}-6\lambda \psi ^2,
\ee
and
\bea\label{peff}
p_{eff}&=&\frac{m_{{\rm eff}}^2}{3}\left(\dot{\psi}+\psi ^2-4H\psi\right)+\kappa ^2f+4\kappa ^2\dot{f}_Q\psi \nonumber\\
&&+\frac{m^2\psi ^2}{2}+6\lambda \psi ^2,
\eea
respectively. In the limiting case $f=0$, $\psi =0$, and $\lambda =\kappa ^2$, the gravitational action (\ref{act}) reduces to the standard Hilbert-Einstein form. In this case $\rho _{eff}=0$, $p_{eff}=0$, and Eqs.~(\ref{gen1}) and (\ref{gen2}) reduce to the standard Friedmann equations of general relativistic cosmology, $3H^2=1/2\kappa ^2\rho$, and $2\dot{H}=-1/2\kappa ^2\left(\rho +p\right)$, respectively.

In order to describe the accelerated/decelerated nature of the cosmological expansion we introduce the deceleration parameter $q$, defined as
\be
q=\frac{d}{dt}\frac{1}{H}-1=-\frac{\dot{H}}{H^2}-1.
\ee
\\

With the use of the generalized Friedmann equations (\ref{gen1}) and (\ref{gen2}) we obtain for the deceleration parameter the expressions
\be
q=\frac{\rho +\rho _{eff}+3\left(p+p_{eff}\right)}{2\left(\rho +\rho _{eff}\right)},
\ee
and
\begin{widetext}
\begin{equation}
q=\frac{\rho +3p+m_{eff}^{2}\left( \dot{\psi}+\psi ^{2}-3H\psi \right)
+2\kappa ^{2}f_{T}\left( \rho +p\right) +2\kappa ^{2}f+12\kappa ^{2}\dot{f}%
_{Q}\psi +m^{2}\psi ^{2}+12\lambda \psi ^{2}}{\rho +m_{eff}^{2}H\psi
+2\kappa ^{2}f_{T}\left( \rho +p\right) -\kappa ^{2}f-m^{2}\psi
^{2}/2-6\lambda \psi ^{2}},
\end{equation}
\end{widetext}
respectively.

\subsubsection{Dimensionless form of the generalized Friedmann equations}

In order to facilitate the comparison of the theoretical results with the cosmological observations,  instead of the usual time variable $t$, we introduce, as independent variable the redshift $z$, defined according to
\be
1 + z =\frac{1}{a},
\ee
where we have used a normalization of the scale factor by imposing the condition that its present
day value is one, $a(0) = 1$. Therefore we can replace the derivatives with respect to the time with the derivatives with respect to the redshift according to the relation
\be
\frac{d}{dt} =\frac{dz}{dt}\frac{d}{dz} = -(1 + z)H(z)\frac{d}{dz} .
\ee
As a function of the cosmological redshift the deceleration parameter $q$ can be obtained as
\be
q(z) = (1 + z)\frac{1}{H(z)}\frac{dH(z)}{dz} - 1.
\ee
In the following we assume that the cosmological matter satisfies the linear barotropic equation of state $p=(\gamma -1)\rho$, where $\gamma $ is a constant, and $1\leq \gamma \leq 2$.

To simplify the mathematical representation of the generalized Friedmann equations we introduce a set of dimensionless variable $\left(\tau, h,r,\Lambda, \Psi,\tilde{Q}\right)$, defined as
\bea
\tau &=& H_0t, H=H_0h, \rho =6\kappa^2H_0^2r, T=6\kappa^2H_0^2\tilde{T}, \nonumber\\
\lambda &=&\kappa^2\Lambda, \psi=H_0\Psi, Q=H_0^2\tilde{Q},f=H_0^2F
\eea
where $H_0$ represents a fixed value of the Hubble function, which may correspond, for example, to the present age of the Universe, or to the end of the inflationary phase of the early Universe. Moreover, $\tilde{Q}=6\Psi ^2$.  Then from equations \eqref{proca}, \eqref{cons}, \eqref{simpfe1} and \eqref{simpfe2}, the equations describing the cosmological evolution in the Weyl type $f(Q,T)$ gravity take the form
\be\label{deqn1}
\frac{d\Psi}{d\tau}=\frac{dh}{d\tau}+2h^2+\Psi ^2-3h\Psi,
\ee
\be\label{deqn2}
\frac{d\Lambda}{d\tau}=-\left(\frac{M^2}{6}+2F_{\tilde{Q}}+2\Lambda\right)\Psi=-\frac{1}{6}M_{eff}^2\Psi,
\ee
\bea\label{deqn3}
\frac{dh}{d\tau}&=&\frac{1}{1-M_{eff}^2/12\Lambda}\Bigg[-\frac{\gamma}{2}\left(3+F_{\bar{T}}\right)\frac{r}{\Lambda}+\frac{\Psi}{\Lambda}\frac{dF_{\tilde{Q}}}{d\tau}\nonumber\\
&&+\frac{M_{eff}^2}{6\Lambda}\left(h^2+\Psi ^2-2h\Psi\right)\Bigg],
\eea
\begin{align}\label{deqn4}
r=\frac{1}{2\left(3+\gamma F_{\tilde{T}}\right)}&\Bigg[F+\frac{M^2\Psi^2}{2}\nonumber\\&+6\Lambda \left(h^2+\Psi^2\right)
-M_{eff}^2h\Psi\Bigg],
\end{align}
where we have denoted
\be\label{meff}
M_{eff}^2=M^2+12F_{\tilde{Q}}+12\Lambda,
\ee
with $M^2=m^2/\emph{\emph{}}\kappa ^2$.

\subsection{The de Sitter solution}

Before we consider more complicated cosmological models, we first investigate a simpler problem, namely, the existence of a de Sitter type vacuum solution of the cosmological field equations. The de Sitter solution corresponds to $\rho=p=0$ and $H=H_0=\text{constant}$, respectively, implying $r=0$, and $h={\rm constant}$, respectively. In this case
Eq.~(\ref{deqn1}) can be solved exactly for $\Psi(t)$, giving
\begin{equation}\label{83}
\tilde{Q} \left( \tau \right) =6h^2\left[ 1+\frac{h-\Psi _{0}}{\left( \Psi
_{0}-2h\right) e^{h\left( \tau -\tau _{0}\right) }+\left( h-\Psi _{0}\right)
}\right]^2 ,
\end{equation}
where $\Psi_0\equiv\Psi(\tau=\tau _0)$.

The simplest possibility of obtaining a de Sitter type solution that satisfies all the field equations is to assume that $M_{eff}^2=0$. Then Eq.~(\ref{deqn2}) immediately gives $\Lambda =\Lambda _0={\rm constant}$, while from Eq.~(\ref{meff}) we obtain for $F$ the simple differential equation
\be
F_{\tilde{Q}}=-\frac{1}{12}\left(M^2+12\Lambda _0\right),
\ee
with the general solution
\be
F\left(\tilde{Q},\tilde{T}\right)=-\frac{1}{12}\left(M^2+12\Lambda _0\right)H_0^2\tilde{Q}+g(\tilde{T}),
\ee
where $g(\tilde{T})$ is an arbitrary integration function of the trace of the energy-momentum tensor. For this functional form of $F$ Eq.~(\ref{deqn3}) is identically satisfied, while Eq.~(\ref{deqn4}) reduces to $g(\tilde{T})+2\Lambda _0h^2=0$, which implies that the function $g\left(\tilde{T}\right)$ must be a constant. Hence, a time varying scalar non-metricity $Q$ as given by Eq.~(\ref{83}) can trigger a de Sitter type accelerated expansion of the Universe. In the limit of large times $\tau \rightarrow \infty$, $\tilde{Q}\rightarrow 6h^2$, thus becoming a constant.

\section{Particular cosmological models}\label{sect3}

In the present Section we will investigate some specific cosmological models in the Weyl type $f(Q, T )$ gravity theory, models that correspond to different choices of the function $f(Q, T )$, describing the nonminimal coupling between the scalar nonmetricity and matter. We will also perform a comparison of the behavior of the geometric and physical cosmological quantities in the Weyl type $f(Q,T)$ gravity with the standard $\Lambda $CDM model, which is based on the observational discovery of the accelerating expansion of the Universe \cite{1a,1b,1c,1d,1e}.

High precision cosmological data have been obtained from the recent study of the Cosmic Microwave Background Radiation by the Planck satellite \cite{1f,1g, 1h}. In our analysis we will adopt the simplifying assumption that the late Universe contains dust matter only, having negligible thermodynamic pressure. Then  the standard general relativistic energy conservation equation $\dot{\rho}+3H\rho=0$ gives for the variation of matter energy density the expression $\rho= \rho _0/a^3=\rho _0 (1+z)^3$, where $\rho _0$ is the present day matter density. The variation of the Hubble function can be obtained as a function of the scale factor in the form \cite{1f}
\be
H=H_0\sqrt{\left(\Omega _b+\Omega _{DM}\right)a^{-3}+\Omega _{\Lambda}},
\ee
where  $\Omega _b$,  $\Omega _{DM}$,and $\Omega _{\Lambda}$ are the density parameters of the baryonic matter, of the cold (pressureless) dark matter,  and of the dark energy (interpreted as a cosmological constant), respectively. The three density parameters satisfy the important relation $\Omega _b+\Omega _{DM}+\Omega _{\Lambda}=1$, indicating that the geometry of the Universe is flat. In a dimensionless form and as a function of the redshift the Hubble function $H(z)=H_0h(z)$ can be written as
\be\label{63}
h(z)=\sqrt{\left(\Omega _{DM}+\Omega _b\right)\left(1+z\right)^{3}+\Omega _{\Lambda}}.
\ee

The deceleration parameter can be obtained as a function of the redshift in the form
\be
q(z)=\frac{3 (1+z)^3 \left(\Omega _{DM}+\Omega _b\right)}{2 \left[\Omega _{\Lambda}+(1+z)^3
   \left(\Omega _{DM}+\Omega _b\right)\right]}-1.
\ee

In our analysis for the density parameters we will adopt the numerical values $\Omega _{DM}=0.2589$, $\Omega _{b}=0.0486$, and $\Omega _{\Lambda}=0.6911$ \cite{1f}, respectively, obtained from the Planck data.  For the total matter density parameter $\Omega _m=\Omega _{DM}+ \Omega _b$ we find the numerical value $\Omega _m=0.3089$. These numerical values of the cosmological parameters give for the present day value of the deceleration parameter  $q(0)=-0.5381$, indicating an accelerating expansion of the Universe. In the standard $\Lambda$CDM cosmological model the variation of the dimensionless matter density with respect to the redshift is given by the expression $r(z)=\Omega _m(1+z)^3=0.3089(1+z)^3$.

In the following we will restrict our investigations to the case of a dust Universe with $\gamma =1$.

\subsection{$f(Q,T)=\alpha Q+\frac{\beta }{6\kappa ^2}T$}

As a first example of a cosmological model in the Weyl type $f(Q,T)$ gravity we will consider the case in which the function $f(Q,T)$ can be represented as
\be
f(Q,T)=\alpha Q+\frac{\beta }{6\kappa ^2}T,
\ee
where $\alpha $ and $\beta $ are constants. After rescaling the variables we obtain for the dimensionless function $F\left(\tilde{Q},\tilde{T}\right)$ the expression  $F\left(\tilde{Q},\tilde{T}\right)=\alpha \tilde{Q}+\beta \tilde{T}$. Hence $F_{\tilde{Q}}=\alpha$, and $F_{\tilde{T}}=\beta$, respectively. For this form of the coupling function the gravitational field equations (\ref{deqn1})-(\ref{deqn4}) take the form
\bea\label{mod1a}
-(1+z)h(z)\frac{d\Psi (z)}{dz}&=&-(1+z)h(z)\frac{dh}{dz}+2h^2(z)\nonumber\\
&&+\psi ^2(z)-3h(z)\psi (z),
\eea
\be\label{mod1b}
(1+z)h(z)\frac{d\Lambda (z)}{dz}=\frac{1}{6}M_{eff}^2(z)\Psi (z),
\ee
\bea\label{mod1c}
&&-(1+z)h(z)\frac{dh(z)}{dz}=\frac{1}{1-M_{eff}^2(z)/12\Lambda (z)}\times \nonumber\\
&&\Bigg\{-\frac{\gamma}{2}\left(3+\beta\right)\frac{r(z)}{\Lambda (z)}
+\frac{M_{eff}(z)^2}{6\Lambda (z)}\Bigg[h^2(z)+\Psi^2(z)\nonumber\\
&&-2h(z)\Psi(z)\Bigg]\Bigg\},
\eea
with
\be
M_{eff}^2(z)=M^2+12\alpha +12\Lambda (z),
\ee
and
\be
r(z)=\frac{\Psi (z) \left[\Psi (z)-2 h(z)\right] M_{eff}^2(z)}{2 (\beta
   +6)}+\frac{6 h^2(z) \Lambda (z)}{\beta +6},
\ee
respectively. The system of differential equations Eqs.~(\ref{mod1a})-(\ref{mod1c}) must be integrated with the initial conditions $h(0)=1$, $\Psi (0)=\Psi _0$, and $\Lambda (0)=\Lambda _0$. The model depends on three free parameters $M^2$ (the mass of the Weyl field), and $\alpha $ and $\beta$, respectively, indicating the strengths of the Weyl geometry-matter coupling.

\begin{figure}[tbp]
\begin{center}
\includegraphics[scale=0.7]{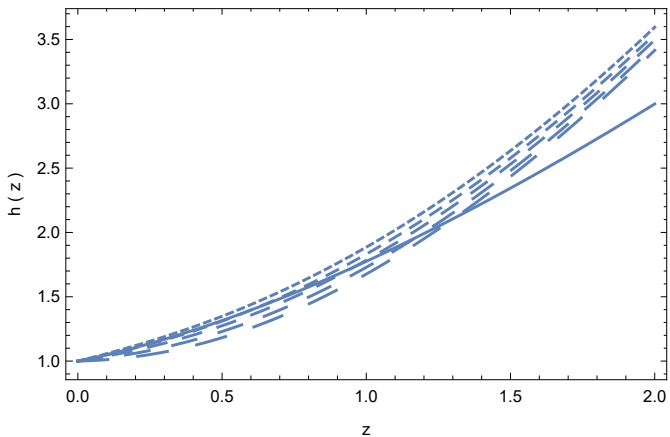}
\caption{Variation as a function of the redshift of the dimensionless Hubble function $h(z)$ for the geometry-matter coupling function $F\left(\tilde{Q},\tilde{T}\right)=\alpha \tilde{Q}+\beta \tilde{T}$ for $\alpha =-1.95$, $M=0.95$, and for different values of $\beta$: $\beta =-2.64$ (dotted curve), $\beta =-2.78$ (short dashed curve), $\beta =-2.91$ (dashed curve), $\beta =-3.07$ (long dashed curve), and $\beta =-3.24$ (ultra-long dashed curve). To integrate the system of cosmological evolution equations we have used the initial conditions $h(0)=1$, $\Psi (0)=0.555$, and $\Lambda (0)=0.568$. The solid curve represents the evolution of the Hubble function in the standard $\Lambda$CDM cosmological model.}\label{fig1}
\end{center}
\end{figure}

\begin{figure}[tbp]
\begin{center}
\includegraphics[scale=0.7]{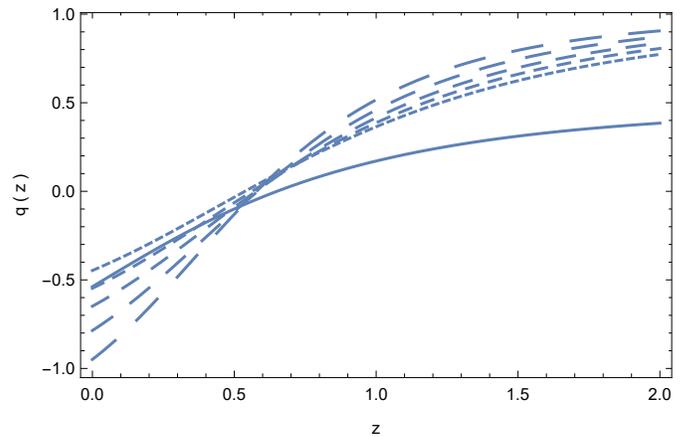}
\caption{Variation as a function of the redshift of the deceleration parameter $q(z)$ for the geometry-matter coupling function $F\left(\tilde{Q},\tilde{T}\right)=\alpha \tilde{Q}+\beta \tilde{T}$ for $\alpha =-1.95$, $M=0.95$, and for different values of $\beta$: $\beta =-2.64$ (dotted curve), $\beta =-2.78$ (short dashed curve), $\beta =-2.91$ (dashed curve), $\beta =-3.07$ (long dashed curve), and $\beta =-3.24$ (ultra-long dashed curve). To integrate the system of cosmological evolution equations we have used the initial conditions $h(0)=1$, $\Psi (0)=0.555$, and $\Lambda (0)=0.568$. The solid curve represents the evolution of the deceleration parameter in the standard $\Lambda$CDM cosmological model.}\label{fig2}
\end{center}
\end{figure}

\begin{figure}[tbp]
\begin{center}
\includegraphics[scale=0.7]{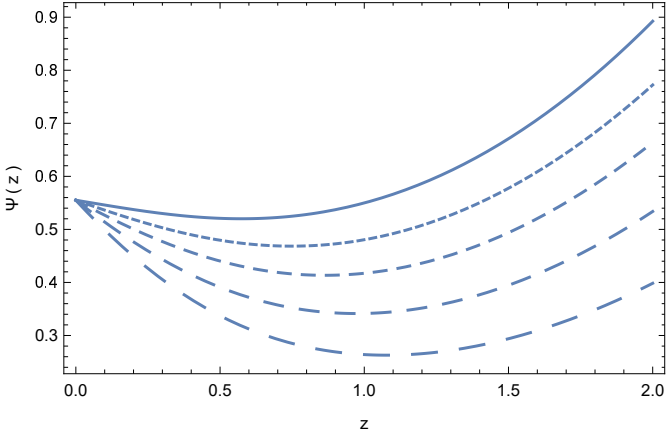}
\caption{Variation as a function of the redshift of the Weyl vector component $\Psi(z)$ for the geometry-matter coupling function $F\left(\tilde{Q},\tilde{T}\right)=\alpha \tilde{Q}+\beta \tilde{T}$ for $\alpha =-1.95$, $M=0.95$, and for different values of $\beta$: $\beta =-2.64$ (solid curve), $\beta =-2.78$ (dotted curve), $\beta =-2.91$ (short dashed curve), $\beta =-3.07$ (dashed curve), and $\beta =-3.24$ (long dashed curve). To integrate the system of cosmological evolution equations we have used the initial conditions $h(0)=1$, $\Psi (0)=0.555$, and $\Lambda (0)=0.568$. }\label{fig4}
\end{center}
\end{figure}

\begin{figure}[tbp]
\begin{center}
\includegraphics[scale=0.7]{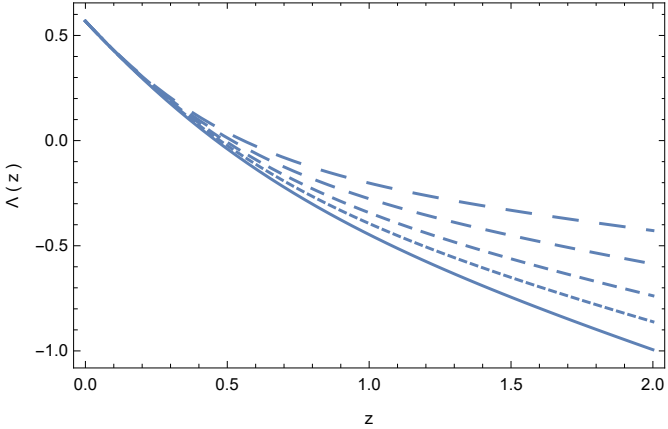}
\caption{Variation as a function of the redshift of the Lagrange multiplier  $\Lambda (z)$ for the geometry-matter coupling function $F\left(\tilde{Q},\tilde{T}\right)=\alpha \tilde{Q}+\beta \tilde{T}$ for $\alpha =-1.95$, $M=0.95$, and for different values of $\beta$: $\beta =-2.64$ (solid curve), $\beta =-2.78$ (dotted curve), $\beta =-2.91$ (short dashed curve), $\beta =-3.07$ (dashed curve), and $\beta =-3.24$ (long dashed curve). To integrate the system of cosmological evolution equations we have used the initial conditions $h(0)=1$, $\Psi (0)=0.555$, and $\Lambda (0)=0.568$.}\label{fig5}
\end{center}
\end{figure}

The variations as a function of redshift of the Hubble function, of the deceleration parameter, of the Weyl vector field, and of the Lagrangian multiplier are represented in Figs.~\ref{fig1}-\ref{fig5}. To obtain the figures we have fixed the numerical value of the dimensionless mass $M^2$ of the Weyl field, and the initial conditions for the Weyl vector $\Psi$, and of the Lagrangian multiplier $\Lambda$. Moreover, we have fixed the value of the parameter $\alpha$, and varied only the parameter $\beta$ in the geometry-matter coupling function $F$.

As one can see from Fig.~\ref{fig1}. the Hubble function is a monotonically increasing function of the redshift (a monotonically decreasing function of time). The evolution of $h(z)$ is strongly dependent on the model parameters, as well as of the initial conditions for $\Psi$ and $\Lambda$. For the chosen set of parameters the model is basically equivalent with the standard $\Lambda$CDM model for a redshift range of the order of $z\in(0,1.25)$.  However, at higher redshifts $z>1.25$ significant differences between the behavior of the Hubble function in the Weyl type $f(Q,T)$ gravity and in the standard $\Lambda$CDM model do appear, with the $f(Q,T)$ model having much higher numerical values, corresponding to a faster expansion rate.

The deceleration parameter $q(z)$, represented in Fig.~\ref{fig2}, shows also a significant dependence on the numerical values of the model parameters. While in the redshift range $z\in (0,0.5)$ the model can reproduce well the results of the standard $\Lambda$CDM model, for higher redshifts the deceleration parameter takes much larger positive values as compared to the $\Lambda$CDM case, indicating a decelerating evolution followed by a quicker transition to the accelerating phase. However, both models enter in the accelerating phase with $q<0$ at the same redshift $z\approx 0.5$, a numerical value that is roughly independent on the numerical values of the considered particular Weyl type $f(Q,T)$ gravity model.


The Weyl vector $\Psi$, whose evolution is depicted in Fig.~\ref{fig4}, shows a complex evolution during the cosmological expansion. For redshifts in the range $z\in (0,1)$, as a function of the redshift the Weyl vector monotonically decreases (increases in time), reaching a minimum value at a redshift of around 1. For $z>1$ the Weyl vector becomes an increasing function of the redshift (a decreasing function of time). The evolution of $\Psi$ is strongly dependent on the model parameters, and a large variety of behaviors are possible. The beginning of the transition towards an accelerating phase of the Universe at a redshift of around $z\approx 1$ is due to the change in the behavior of the Weyl vector, which, after decreasing in time in the early stages of the expansion of the Universe, experiences a transition to an increasing phase, thus triggering the recent cosmological acceleration.

The Lagrange multiplier $\Lambda$, portrayed in Fig.~\ref{fig5}, is a monotonically decreasing function of the redshift, and thus an increasing function of the cosmological time. For low redshifts in the range $(0,0.5)$ the evolution of $\Lambda$ is basically independent on the model parameters. However, at higher redshifts, $\Lambda (z)$ strongly depends on the  numerical values of the model parameters, and for $z>0.5$ it takes negative numerical values.

\subsection{$f(Q,T)=\frac{\alpha}{6H_0^2\kappa ^2}QT$}

\begin{figure}[tbp]
\begin{center}
\includegraphics[scale=0.7]{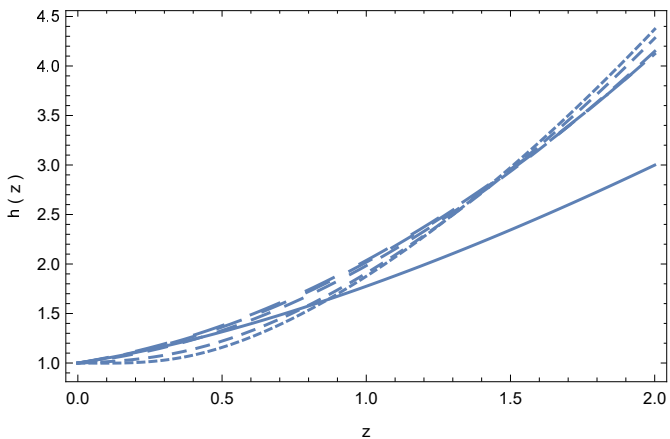}
\caption{Variation as a function of the redshift of the dimensionless Hubble function $h(z)$ for the geometry-matter coupling function $F\left(\tilde{Q},\tilde{T}\right)=\alpha \tilde{Q}\tilde{T}$ for $M=0.98$, and for different values of $\alpha $: $\alpha =-0.72$ (dotted curve), $\alpha =-0.77$ (short dashed curve), $\alpha =-0.80$ (dashed curve), $\alpha =-0.81$ (long dashed curve), and $\alpha =-0.82$ (ultra-long dashed curve). To integrate the system of cosmological evolution equations we have used the initial conditions $h(0)=1$, $\Psi (0)=0.455$, and $\Lambda (0)=0.335$. The solid curve represents the evolution of the Hubble function in the standard $\Lambda$CDM cosmological model.}\label{fig6}
\end{center}
\end{figure}

\begin{figure}[tbp]
\begin{center}
\includegraphics[scale=0.7]{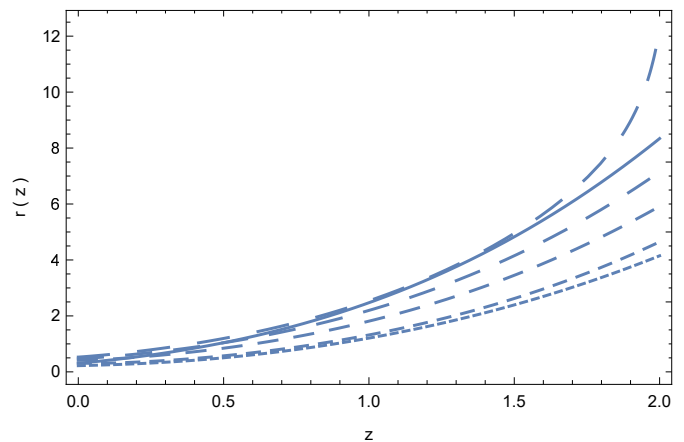}
\caption{Variation as a function of the redshift of the matter energy density $r(z)$ for the geometry-matter coupling function $F\left(\tilde{Q},\tilde{T}\right)=\alpha \tilde{Q}\tilde{T}$ for $M=0.98$, and for different values of $\alpha $: $\alpha =-0.72$ (dotted curve), $\alpha =-0.77$ (short dashed curve), $\alpha =-0.80$ (dashed curve), $\alpha =-0.81$ (long dashed curve), and $\alpha =-0.82$ (ultra-long dashed curve). To integrate the system of cosmological evolution equations we have used the initial conditions $h(0)=1$, $\Psi (0)=0.455$, and $\Lambda (0)=0.335$. The solid curve represents the evolution of the matter density in the standard $\Lambda$CDM cosmological model.}\label{fig7}
\end{center}
\end{figure}

\begin{figure}[tbp]
\begin{center}
\includegraphics[scale=0.7]{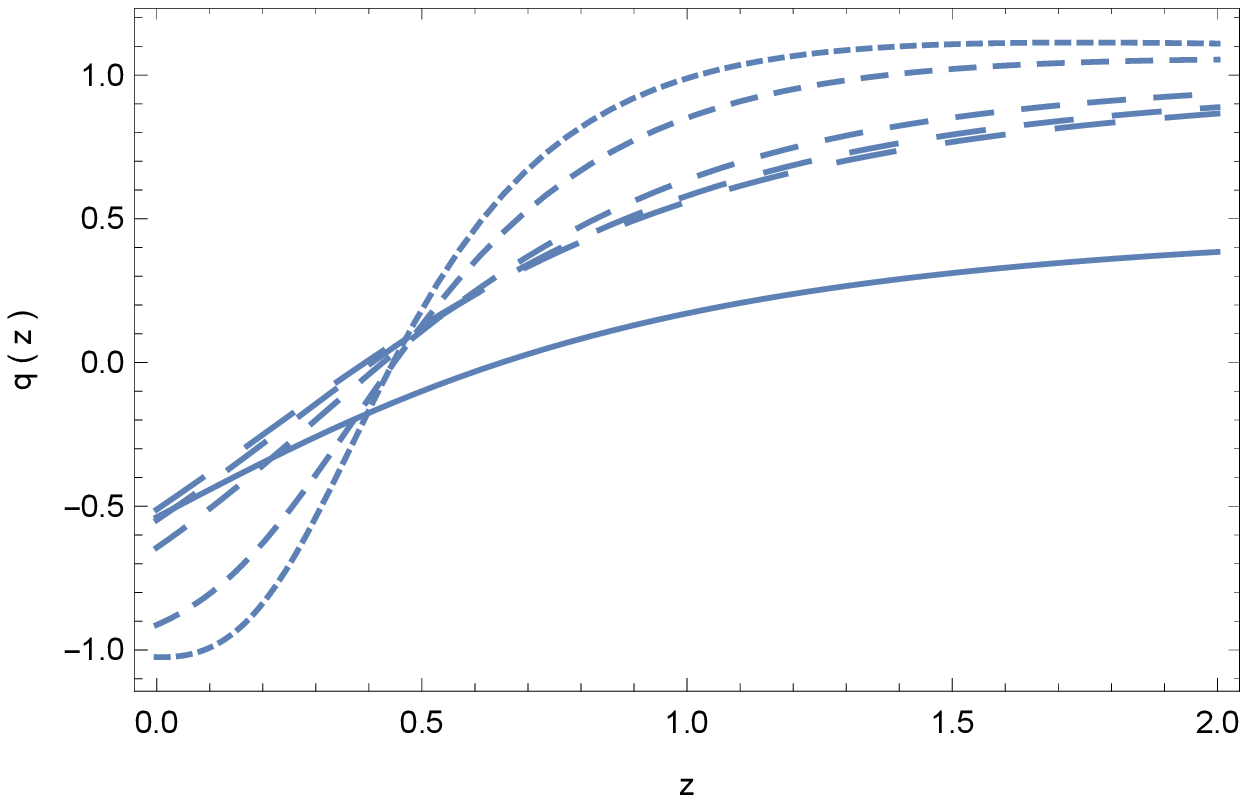}
\caption{Variation as a function of the redshift of the deceleration parameter $q\emph{}(z)$ for the geometry-matter coupling function $F\left(\tilde{Q},\tilde{T}\right)=\alpha \tilde{Q}\tilde{T}$ for $M=0.98$, and for different values of $\alpha $: $\alpha =-0.72$ (dotted curve), $\alpha =-0.77$ (short dashed curve), $\alpha =-0.80$ (dashed curve), $\alpha =-0.81$ (long dashed curve), and $\alpha =-0.82$ (ultra-long dashed curve). To integrate the system of cosmological evolution equations we have used the initial conditions $h(0)=1$, $\Psi (0)=0.455$, and $\Lambda (0)=0.335$. The solid curve represents the evolution of the deceleration parameter in the standard $\Lambda$CDM cosmological model.}\label{fig8}
\end{center}
\end{figure}

\begin{figure}[tbp]
\begin{center}
\includegraphics[scale=0.7]{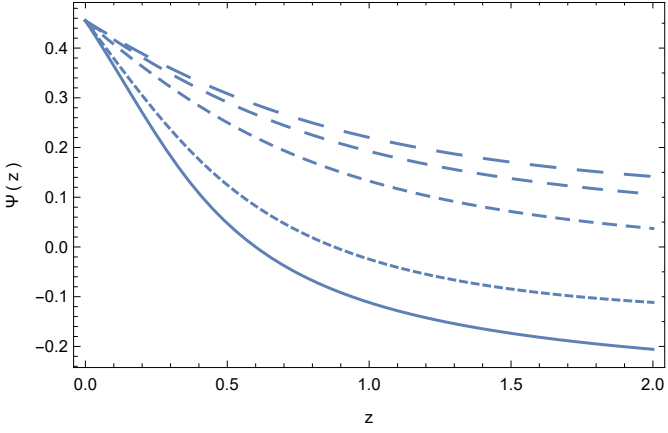}
\caption{Variation as a function of the redshift of the Weyl vector component $\Psi(z)$ for the geometry-matter coupling function $F\left(\tilde{Q},\tilde{T}\right)=\alpha \tilde{Q}\tilde{T}$ for $M=0.98$, and for different values of $\alpha $: $\alpha =-0.72$ (dotted curve), $\alpha =-0.77$ (short dashed curve), $\alpha =-0.80$ (dashed curve), $\alpha =-0.81$ (long dashed curve), and $\alpha =-0.82$ (ultra-long dashed curve). To integrate the system of cosmological evolution equations we have used the initial conditions $h(0)=1$, $\Psi (0)=0.455$, and $\Lambda (0)=0.335$. }\label{fig9}
\end{center}
\end{figure}

\begin{figure}[tbp]
\begin{center}
\includegraphics[scale=0.7]{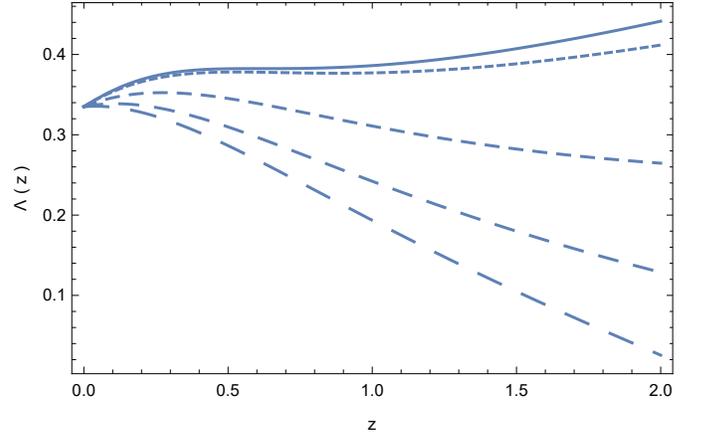}
\caption{Variation as a function of the redshift of the Lagrange multiplier  $\Lambda (z)$ for the geometry-matter coupling function $F\left(\tilde{Q},\tilde{T}\right)=\alpha \tilde{Q}\tilde{T}$ for $M=0.98$, and for different values of $\alpha $: $\alpha =-0.72$ (dotted curve), $\alpha =-0.77$ (short dashed curve), $\alpha =-0.80$ (dashed curve), $\alpha =-0.81$ (long dashed curve), and $\alpha =-0.82$ (ultra-long dashed curve). To integrate the system of cosmological evolution equations we have used the initial conditions $h(0)=1$, $\Psi (0)=0.455$, and $\Lambda (0)=0.335$.}\label{fig10}
\end{center}
\end{figure}

As a second example of a cosmological model in the Weyl type $f(Q,T)$ gravity we assume that the function $f(Q,T)$ can be represented as $f(Q,T)=\left(\alpha/6H_0^2\kappa ^2\right)QT$, where $\alpha $ is a constant. Then we obtain successively $F\left(\tilde{Q},\tilde{T}\right)=\alpha \tilde{Q}\tilde{T}$, $F_{\tilde{T}}=\alpha \tilde{Q}=6 \Psi ^2$, and $f_{\tilde{Q}}-\alpha \tilde{T}=\alpha r$, respectively. The system of differential equations describing the cosmological evolution in this model takes the form
\bea\label{mod2a}
-(1+z)h(z)\frac{d\Psi (z)}{dz}&=&-(1+z)h(z)\frac{dh}{dz}+2h^2(z)\nonumber\\
&&+\psi ^2(z)-3h(z)\psi (z),
\eea
\be\label{mod2b}
(1+z)h(z)\frac{d\Lambda (z)}{dz}=\frac{1}{6}M_{eff}^2(z)\Psi (z),
\ee
\bea\label{mod2c}
&&-(1+z)h(z)\frac{dh(z)}{dz}=\frac{1}{1-M_{eff}^2(z)/12\Lambda (z)}\times \nonumber\\
&&\Bigg\{-\frac{3\gamma}{2}\left[1+2\alpha \Psi^2(z)\right]\frac{r(z)}{\Lambda (z)}
-\alpha (1+z)h(z)\frac{\Psi (z)}{\Lambda (z)}\frac{dr(z)}{dz}\nonumber\\
&&+\frac{M_{eff}(z)^2}{6\Lambda (z)}\Bigg[h^2(z)+\Psi^2(z)-2h(z)\Psi(z)\Bigg]\Bigg\},
\eea
where
\be
M_{eff}^2(z)=M^2+12\alpha r (z) +12\Lambda (z),
\ee
and
\begin{eqnarray}
r(z) &=&\frac{1}{12\left\{ \alpha \Psi (z)\left[ 2h(z)+(2\gamma -1)\Psi (z)%
\right] +1\right\} }\times  \nonumber\\
&&\Bigg\{ 2h(z)\Psi (z)\left[ M^{2}+12\Lambda (z)\right] +12h^{2}(z)\Lambda
(z)\nonumber\\
&&+\Psi ^{2}(z)\left[ M^{2}+12\Lambda (z)\right] \Bigg\} ,
\end{eqnarray}
respectively. The system of strongly nonlinear system of differential equations (\ref{mod2a})-(\ref{mod2c}) must be integrated with the initial conditions $h(0)=1$, and $\Psi (0)=\Psi_0$ and $\Lambda (0)=\Lambda _0$, respectively. The variations of the Hubble function, matter density, deceleration parameter, Weyl vector field, and of the Lagrange multiplier are represented in Figs.~\ref{fig6}-\ref{fig10}.  To obtain the figures we have fixed the initial (present day) values of the Weyl vector and of the Lagrange multiplier, and we have slightly varied the numerical value of $\alpha$.

The variation of the Hubble function in this model is represented in Fig.~\ref{fig6}. In the considered redshift range the Hubble function is an increasing function of $z$ (a decreasing function of time). Up to a redshift $z\approx 1$ the model can give a good alternative description of the standard $\Lambda$CDM model. For higher redshifts the differences between the predictions of the Weyl type $f(Q,T)$ gravity and the standard cosmological model become significant, with the Weyl type $f(Q,T)$ gravity predicting, at least for the chosen set of parameters, higher numerical values, and a more significant increase with respect to $z$ of $h(z)$. The variation of $h(z)$ is relatively independent of the variation of the values of $\alpha$.

The variation with the redshift of the matter energy density $r(z)$ is represented in Fig.~\ref{fig7}. The energy density is an increasing function of $z$, or, equivalently, a decreasing function of the cosmological time.  As one can see from the Figure, the Weyl type $f(Q,T)\propto QT$ model can give a good alternative description of the matter dynamics, with some values of the model parameter $\alpha$ reproducing almost exactly the standard $\Lambda$CDM model up to the redshift $z\approx 1.5$. However, the matter density is strongly dependent on the choice of the model parameter $\alpha$, and a large number of evolutionary scenarios for matter can be constructed by varying the values of $\alpha$.

The deceleration parameter of the model, depicted in Fig.~\ref{fig8}, is also strongly dependent on the numerical values of $\alpha$, and thus allows the possibility of constructing a large number of cosmological evolutionary expansions. For $z<0.5$ we obtain a good qualitative concordance with the predictions of the standard cosmology. Cosmological phases with a de Sitter type expansion with $q=-1$ can also be obtained. However, the nature of the transition to the accelerating phase is different in the Weyl and standard cosmology. Up to redshifts of around $z\approx 1$, the deceleration parameter is roughly a constant in the range $q\in (0.5,1)$, and the Universe is decelerating. Then the Universe began to accelerate, and after a short cosmological time interval it entered in an accelerating phase, with $q<0$. The value of $\alpha $ determines the nature of the final stages of the accelerating evolution.

The Weyl vector, whose behavior is indicated in Fig.~\ref{fig9}, is a monotonically decreasing function of $z$, and a monotonically increasing function of the cosmological time. The temporal increase of $\Psi$ triggers the late acceleration of the Universe, with the rate of change of $\Psi$ becoming significant after the redshift $z\approx 1$. The evolution of $\Psi$ is strongly dependent on the numerical values of the model parameter $\alpha$, as well as of the initial conditions adopted to integrate the cosmological evolution equation.

The evolution of the Lagrange multiplier $\Lambda (z)$, represented in Fig.~\ref{fig10}, is strongly dependent on the numerical values of $\alpha$, and indicates a complex behavior. For larger values of $\alpha$, $\Lambda (z)$ shows an oscillating behavior, with increasing phases alternating with decreasing ones. For smaller values of $\alpha$ the Lagrange multiplier is a monotonically decreasing function of $z$, indicating its increase in time. In the present model the combined effects of $\Psi$ and $\Lambda$ at low redshifts determine the transition of the Universe from a decelerating to an accelerating phase.

\subsection{$f(Q,T)=\eta H_0^2e^{\frac{\mu}{6H_0^2} Q}+\frac{\nu}{6\kappa ^2} T$}

\begin{figure}[tbp]
\begin{center}
\includegraphics[scale=0.7]{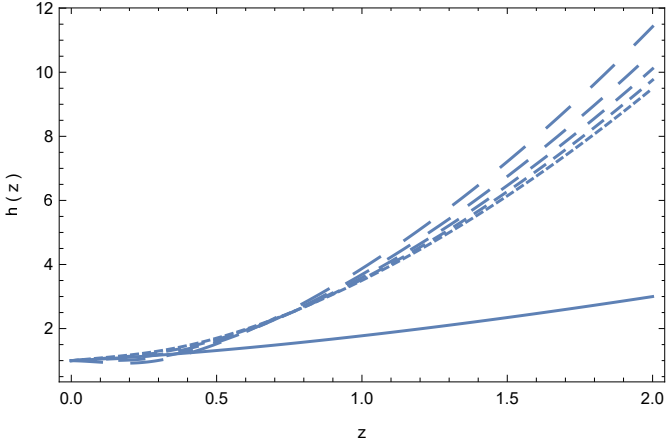}
\caption{Variation as a function of the redshift of the dimensionless Hubble function $h(z)$ for the geometry-matter coupling function $F\left(\tilde{Q},\tilde{T}\right)=\eta e^{\mu \tilde{Q}/6}+\nu \tilde{T}$ for $M=1.7$, $\eta =6/\mu$, $\nu =20$, and for different values of $\mu $: $\mu =1.7$ (dotted curve), $\mu= =1.5$ (short dashed curve), $\mu =1.3$ (dashed curve), $\mu =1.1$ (long dashed curve), and $\mu =0.9$ (ultra-long dashed curve). To integrate the system of cosmological evolution equations we have used the initial conditions $h(0)=1$, $\Psi (0)=0.058$, and $\Lambda (0)=0.0235$. The solid curve represents the evolution of the Hubble function in the standard $\Lambda$CDM cosmological model.}\label{fig11}
\end{center}
\end{figure}

\begin{figure}[tbp]
\begin{center}
\includegraphics[scale=0.7]{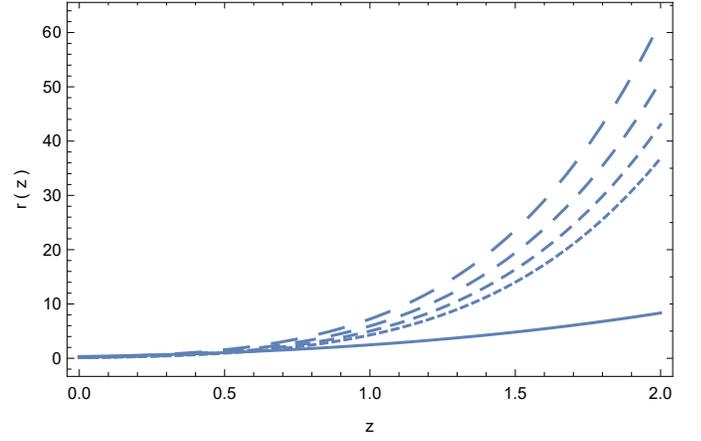}
\caption{Variation as a function of the redshift of the matter energy density $r(z)$ for the geometry-matter coupling function $F\left(\tilde{Q},\tilde{T}\right)=\eta e^{\mu \tilde{Q}/6}+\nu \tilde{T}$ for $M=1.7$, $\eta =6/\mu$, $\nu =20$, and for different values of $\mu $: $\mu =1.7$ (dotted curve), $\mu= =1.5$ (short dashed curve), $\mu =1.3$ (dashed curve), $\mu =1.1$ (long dashed curve), and $\mu =0.9$ (ultra-long dashed curve). To integrate the system of cosmological evolution equations we have used the initial conditions $h(0)=1$, $\Psi (0)=0.058$, and $\Lambda (0)=0.0235$. The solid curve represents the evolution of the matter density in the standard $\Lambda$CDM cosmological model.}\label{fig12}
\end{center}
\end{figure}

\begin{figure}[tbp]
\begin{center}
\includegraphics[scale=0.7]{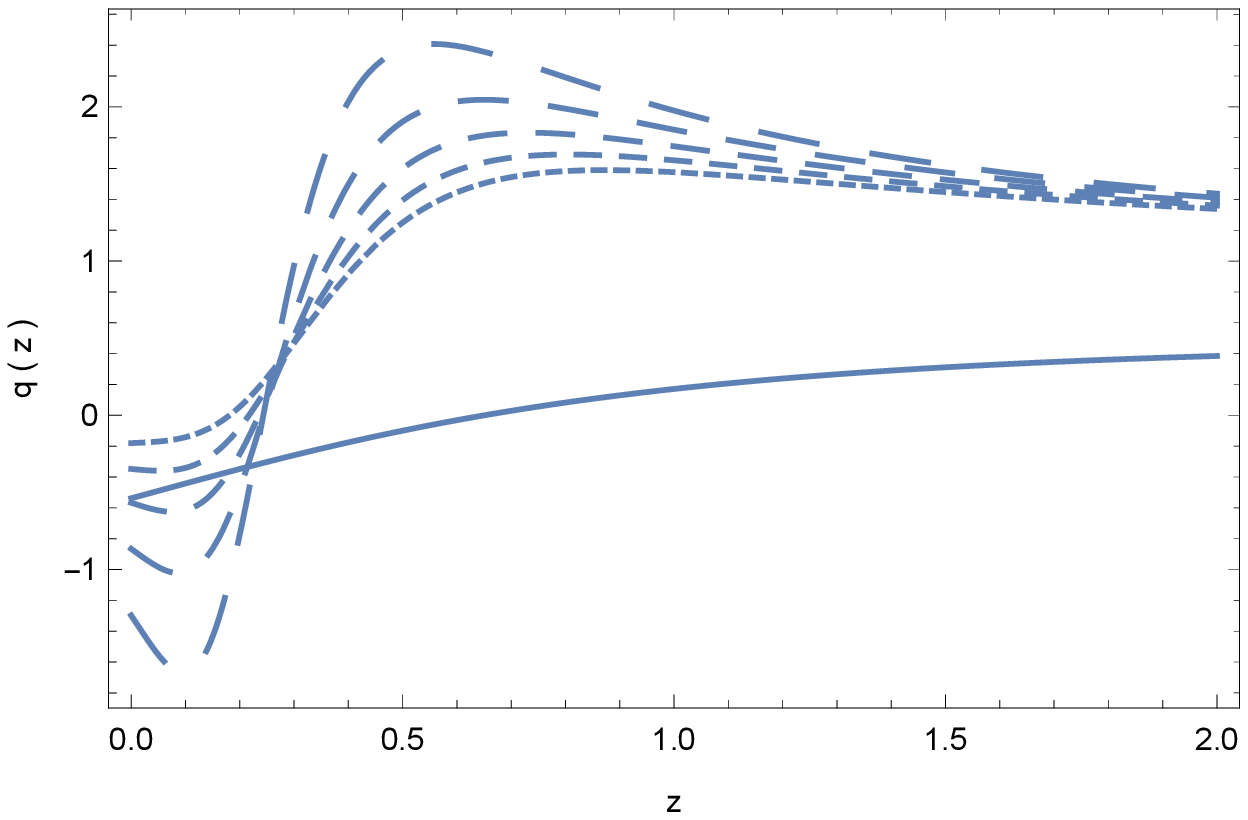}
\caption{Variation as a function of the redshift of the deceleration parameter $q\emph{}(z)$ for the geometry-matter coupling function $F\left(\tilde{Q},\tilde{T}\right)=\eta e^{\mu \tilde{Q}/6}+\nu \tilde{T}$ for $M=1.7$, $\eta =6/\mu$, $\nu =20$, and for different values of $\mu $: $\mu =1.7$ (dotted curve), $\mu= =1.5$ (short dashed curve), $\mu =1.3$ (dashed curve), $\mu =1.1$ (long dashed curve), and $\mu =0.9$ (ultra-long dashed curve). To integrate the system of cosmological evolution equations we have used the initial conditions $h(0)=1$, $\Psi (0)=0.058$, and $\Lambda (0)=0.0235$. The solid curve represents the evolution of the deceleration parameter in the standard $\Lambda$CDM cosmological model.}\label{fig13}
\end{center}
\end{figure}

\begin{figure}[tbp]
\begin{center}
\includegraphics[scale=0.7]{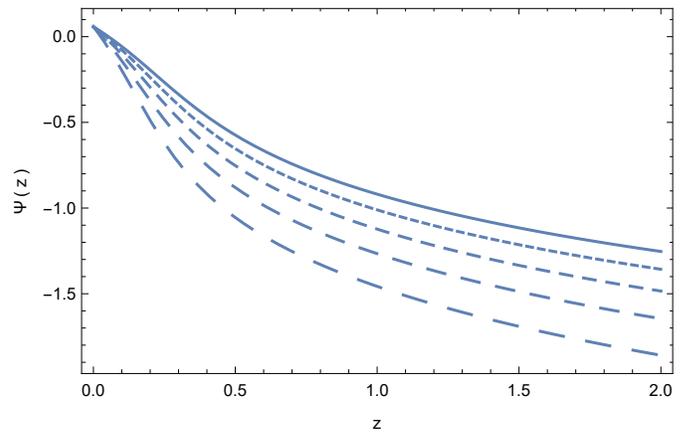}
\caption{Variation as a function of the redshift of the Weyl vector component $\Psi(z)$ for the geometry-matter coupling function $F\left(\tilde{Q},\tilde{T}\right)=\eta e^{\mu \tilde{Q}/6}+\nu \tilde{T}$ for $M=1.7$, $\eta =6/\mu$, $\nu =20$, and for different values of $\mu $: $\mu =1.7$ (dotted curve), $\mu= =1.5$ (short dashed curve), $\mu =1.3$ (dashed curve), $\mu =1.1$ (long dashed curve), and $\mu =0.9$ (ultra-long dashed curve). To integrate the system of cosmological evolution equations we have used the initial conditions $h(0)=1$, $\Psi (0)=0.058$, and $\Lambda (0)=0.0235$. }\label{fig14}
\end{center}
\end{figure}

\begin{figure}[tbp]
\begin{center}
\includegraphics[scale=0.7]{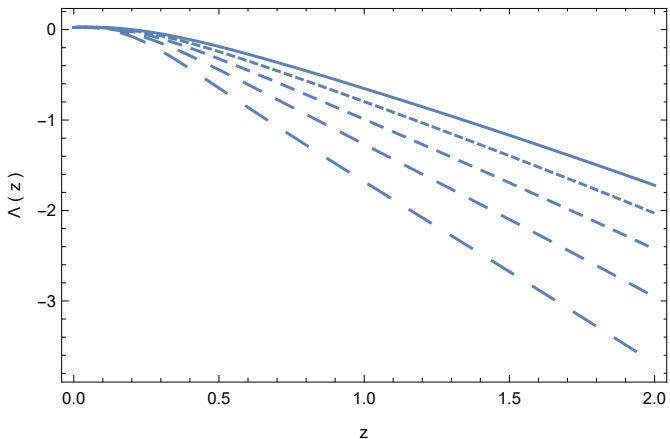}
\caption{Variation as a function of the redshift of the Lagrange multiplier  $\Lambda(z)$ for the geometry-matter coupling function $F\left(\tilde{Q},\tilde{T}\right)=\eta e^{\mu \tilde{Q}/6}+\nu \tilde{T}$ for $M=1.7$, $\eta =6/\mu$, $\nu =20$, and for different values of $\mu $: $\mu =1.7$ (dotted curve), $\mu= =1.5$ (short dashed curve), $\mu =1.3$ (dashed curve), $\mu =1.1$ (long dashed curve), and $\mu =0.9$ (ultra-long dashed curve). To integrate the system of cosmological evolution equations we have used the initial conditions $h(0)=1$, $\Psi (0)=0.058$, and $\Lambda (0)=0.0235$.}\label{fig15}
\end{center}
\end{figure}

As a third possible cosmological model in the Weyl type $f(Q,T)$ gravity we will consider the case in which the function $f(Q,T)$ is given by $f(Q,T)=\eta H_0^2e^{\left(\mu /6H_0^2\right)Q}+\left(\nu /6\kappa ^2\right)T$, where $\eta $, $\mu$ and $\nu $ are constants. Hence we obtain $F\left(\tilde{Q},\tilde{T}\right)=\eta e^{\mu \tilde{Q}/6}+\nu r$, $F_{\tilde{Q}}=(\eta \mu /6)e^{\mu \tilde{Q}/6}$, and $F_{\tilde{T}}=\nu$, respectively. In the following for simplicity we assume that $\eta \mu /6=1$. For this choice of the function $f(Q,T)$  the cosmological evolution equations take the form
\bea\label{mod3a}
-(1+z)h(z)\frac{d\Psi (z)}{dz}&=&-(1+z)h(z)\frac{dh}{dz}+2h^2(z)\nonumber\\
&&+\psi ^2(z)-3h(z)\psi (z),
\eea
\be\label{mod3b}
(1+z)h(z)\frac{d\Lambda (z)}{dz}=\frac{1}{6}M_{eff}^2(z)\Psi (z),
\ee
\bea\label{mod3c}
&&-(1+z)h(z)\frac{dh(z)}{dz}=\frac{1}{1-M_{eff}^2(z)/12\Lambda (z)}\times \nonumber\\
&&\Bigg\{-\frac{\gamma}{2}\left(3+\nu\right)\frac{r(z)}{\Lambda (z)}
-\mu (1+z)h(z)\frac{\Psi ^2(z)}{\Lambda (z)}\frac{d\Psi(z)}{dz}\times \nonumber\\
&&e^{\mu \Psi ^2(z)}+\frac{M_{eff}(z)^2}{6\Lambda (z)}\Bigg[h^2(z)+\Psi^2(z)-2h(z)\Psi(z)\Bigg]\Bigg\},\nonumber\\
\eea
where
\be
M_{eff}^2(z)=M^2+12e^{ \mu \Psi^2 (z)} +12\Lambda (z),
\ee
and
\begin{eqnarray}
\hspace{-0.7cm}&&r(z) =-\frac{h(z) \Psi (z) \left[M^2+12 \Lambda (z)\right]}{\nu +6}+\frac{6 h^2(z) \Lambda
   (z)}{\nu +6}\nonumber\\
 \hspace{-0.7cm} && -\frac{6 e^{\mu  \Psi ^2(z)} \left[2 \mu  h(z) \Psi (z)-1\right]}{\mu  (\nu
   +6)}+\frac{\Psi ^2(z) \left[M^2+12 \Lambda (z)\right]}{2 (\nu +6)},\nonumber\\
\end{eqnarray}
respectively. The system of equations (\ref{mod3a})-(\ref{mod3c}) must be integrated with the initial conditions $h(0)=1$, $\Psi (0)=\Psi _0$ and $\Lambda (0)=\Lambda _0$, respectively. The variations with the redshift of the Hubble function $h(z)$, matter energy density $r(z)$, deceleration parameter $q(z)$, Weyl vector $\Psi (z)$, and of the Lagrange multiplier $\Lambda (z)$ are represented in Figs.~\ref{fig11}-\ref{fig15}, respectively.

The variation of the Hubble function in this model is represented in Fig.~\ref{fig11}. The Hubble function of this model is an increasing function of $z$ (a decreasing function of time), indicating an expansionary evolution of the Universe. Up to a redshift $z\approx 0.5$ the model reproduces well the standard $\Lambda$CDM model, and for the redshift range $z\in (0,0.5)$, the cosmological evolution does not depend significantly on the numerical values of the parameters. For redshifts $z>0.5$ the differences between the predictions of the Weyl type $F\left(\tilde{Q},\tilde{T}\right)=\eta e^{\mu \tilde{Q}/6}+\nu \tilde{T}$ gravitational model and the standard cosmological model become important, with the present Weyl type $f(Q,T)$ gravity model predicting, for the chosen set of parameters, much higher numerical values, and a rapid increase of $h(z)$ with respect to $z$. At high redshifts the variation of $h(z)$ is strongly dependent on the numerical values of the parameter $\mu$.

The variation with the redshift of the matter energy density $r(z)$ is represented in Fig.~\ref{fig12}. The energy density is an increasing function of $z$, with the matter energy density decreasing in time.  For small redshifts $0\leq <1$ the predictions of the Weyl type $F\left(\tilde{Q},\tilde{T}\right)=\eta e^{\mu \tilde{Q}/6}+\nu \tilde{T}$ gravity reproduce well the standard $\Lambda$CDM cosmological model, while in the redshift range $z<0.5$ the predictions of the two models are very close. In the small redshift range the evolution is independent on the model parameters. At larger redshifts  the matter density becomes strongly dependent on the model parameters, and significant differences do appear as compared to the standard cosmology, with the matter density taking much higher values.

The deceleration parameter of the model, represented in Fig.~\ref{fig13}, shows important differences with respect to the $\Lambda$CDM model. If for small redshifts one could find a set of parameters that reproduce relatively well standard cosmology, at higher redshifts both the qualitative and quantitative differences become important. The Universe still strongly decelerates at redshifts $z>0.5$, with the deceleration parameter taking values of $q\approx 2$. At a redshift of around $z\approx 0.5$, the Universe experiences a transition to an accelerating phase, and after a short cosmological interval the deceleration parameter takes negative values. The final evolutionary stages are strongly dependent on the model parameters, and the Universe can enter in a de Sitter phase with $q\approx -1$. Cosmological scenarios with $q<-1$ can also be obtained within this model.

The Weyl vector, presented in Fig.~\ref{fig14}, is a monotonically decreasing function of $z$, and a monotonically increasing function of time. It takes negative values, except for a small redshift range near the origin $z=0$. The late acceleration of the Universe is determined by the increase of $\Psi$. For high redshifts the evolution of $\Psi$ depends significantly on the numerical values of the model parameters, and on the initial conditions used to numerically integrate the cosmological evolution equation.

The Lagrange multiplier $\Lambda (z)$, shown in Fig.~\ref{fig15}, is also a monotonically decreasing function of the redshift. Except for a small region near the origin $z=0$ it takes negative values. For higher redshifts the evolution of $\Lambda (z)$ depends strongly on the model parameters.

\section{Discussions and final remarks}\label{sect4}

Despite its initial success in describing the cosmological observations, presently several disagreements between the $\Lambda$CDM model and observations raise the possibility that in fact standard cosmology is just an approximation of a more realistic theory. Beyond the "standard" problems of dark matter and dark energy, a number of other inconsistencies do appear when confronting the model with astronomical data. For example, the value of the Hubble constant obtained from  the Planck CMB anisotropies is significantly smaller than the values derived by using the luminosity
distances of supernovae \cite{Riess}. Moreover, recent cosmic shear surveys have shown that  the combination of the matter density $\Omega _m$, parameterized by the $S_8 \equiv \sigma _8\sqrt{\Omega _m/0.3}$ parameter, and of the amplitude of the dark matter fluctuations $\sigma_8$ on scales of the order of 8 Mpc $h^{-1}$ is significantly smaller than the value obtained from the Planck data and the assumption of the  $\Lambda$CDM model \cite{Asgari}. In \cite{Silk} it was shown that a combined analysis of the CMB anisotropy power spectra, obtained by the Planck satellite, and luminosity distance data simultaneously excludes a flat Universe and a cosmological constant at 99\% C.L. These results are valid when combining Planck with three different datasets.

From a theoretical point of view we are also witnessing an interesting situation. One of the important moments in the development of theoretical physics happened more than one hundred years since Einstein did propose the first geometric description of gravity, the general relativity theory, we are presently facing the unusual circumstance that at least three geometric descriptions of gravity are possible. The different versions can be constructed independently with the use of the three basic concepts introduced in Riemannian geometry, and its extensions, namely, the curvature, torsion and nonmetricity of the spacetime, respectively. These intriguing developments raise the fundamental question if a unique geometric description of gravity is really possible. Are the three independent descriptions of gravity completely equivalent, or maybe they are only particular cases of a more general geometric theory of gravity, which still needs to be discovered?

In the present paper we have investigated a particular representation of the third geometric description of gravity, represented by the symmetric teleparallel gravity, or $f(Q)$ gravity, in which the basic quantity describing the gravitational field is the nonmetricity $Q$. Moreover, we have considered the class of theories, introduced in \cite{s21}, in which the nonmetricity $Q$ is coupled nonminimally to the trace of the matter energy-momentum tensor $T$. The $f(Q,T)$ theory is constructed in a way similar to the $f(R,T)$ theory \cite{fRT,book}, but with the standard Ricci scalar replaced by the nonmetricity that describes the symmetric teleparallel formulation of gravity. Similarly to the case of the standard curvature - matter couplings, in the $f(Q,T)$ theory the coupling  between $Q$ and $T$ leads to the nonconservation of the energy-momentum tensor. But, in the present approach to $f(Q,T)$ type gravity theories, instead of keeping the nonmetricity $Q$ arbitrary, we have fixed it from the beginning, by using the prescriptions of the Weyl geometry, in which the covariant divergence of the metric tensor is given by the product of the metric and of the Weyl vector $w_{\mu}$. The scalar nonmetricity is related in a simple way to the square of the Weyl vector as $Q=-6w^2$, and thus all the geometric properties of the theory are determined by the Weyl vector and the metric tensor, respectively. To obtain the gravitational field equations of the Weyl type $f(Q,T)$ gravity we have introduced a variational principle, which generalizes the variational principle of the $f(Q,T)$ theory, and whose gravitational sector is constructed from three components. The first component is an arbitrary function of the Weyl vector and of the trace of the matter energy-momentum tensor $f(-6w^2,T)$. The second component is represented by the kinetic term and the mass term of the field, assumed to be massive. Finally, we have adopted the teleparallel view on gravitation and geometry, by assuming that the Ricci-Weyl scalar of the spacetime identically vanishes. This condition is introduced in the gravitational action via a Lagrange multiplier. By varying the gravitational action with respect to the metric and the Weyl vector we have obtained the system of gravitational field equations, which described gravity in terms of the metric and a vector field, and a generalized Proca type equation for the evolution of the Weyl vector. We have performed our analysis of the gravitational action in the framework of the
metric-affine formalism. The covariant divergence of the matter energy-momentum tensor  has also been obtained, and it turns out that generally it is not conserved. The energy and momentum balance equations have been derived explicitly from the matter nonconservation relation. The nonconservation of the matter energy-momentum tensor may have important physical implications, leading to significant changes in the thermodynamics of the Universe, similarly to those in the theories with geometry-matter coupling \cite{fRT,book, Bert1,Bert2,Bert3}. Moreover,  due to the nongeodesic motion of test particles induced by the geometry matter coupling, in the present approach, similarly to other modified gravity theories \cite{Bert1, Bert5} an extra force acting on massive  particles is generated.

The investigations presented in the present paper may also lead to a better understanding of the
geometrical formulation of gravity theories, including the aspects related to the geometry-matter coupling. The present approach allows a  consistent representation of the $f(Q,T)$ type theories with nonminimal curvature-matter coupling.

As a first observational test of the Weyl type $f(Q,T)$ gravity theory we have analyzed its cosmological implications. As a first step in this direction we have obtained the generalized Friedmann equations of the Weyl type $f(Q,T)$ theory. For the description of the Universe we have adopted the  homogeneous and isotropic  Friedmann-Lemaitre-Robertson-Walker type metric, describing the cosmological evolution in a flat geometry. In the Weyl type $f(Q,T)$ theory we can reformulate the generalized Friedmann equations as the standard Friedmann equations of general relativity in which the ordinary matter energy density and pressure are replaced by some effective quantities $\rho _{eff}$ and $p_{eff}$. The effective thermodynamic parameters $\rho _{eff}$ and $p_{eff}$ depend first on the function $f(Q,T)$ and of its derivatives with respect to $Q$ and $T$, which are effectively functions of the Weyl vector and of the thermodynamic parameters of matter. There is also an explicit dependence on the Weyl vector, its time derivative, the mass of the vector field, and on the Lagrange multiplier $\lambda$. The effective thermodynamic energy also contains the linear combination of the ordinary matter energy density and pressure, multiplied by $f_T$. Hence the basic equations describing
the cosmological evolution in the Weyl type $f(Q,T)$ gravity can be formulated in terms of two effective thermodynamic quantities, an energy density and pressure, respectively, which
depend on the matter energy and pressure components of the energy-momentum tensor, on the Weyl vector, and on the Lagrange multiplier, respectively. The function $f(Q,T)$ and its derivatives are effective functions of the Weyl vector and of $T$. Hence in the present model the evolution of the Universe is controlled by the Weyl vector, the Lagrange multiplier, and the matter content. An important indicator of the nature of the cosmological evolution is the deceleration parameter. $q$ can be expressed in terms of the matter energy density and pressure, of the Weyl vector, and of the Lagrange multiplier. The deceleration parameter has a complicated dependence on the function $f$ and of its derivatives. Hence, depending on the functional form of $f(Q,T)$, a large variety of cosmological evolutions can be obtained in the framework of the Weyl type $f(Q,T)$ gravity, including accelerating and decelerating cosmological expansions. For the vacuum case when $\rho =p=0$ we have also shown explicitly that the field equations of Weyl type $f(Q,T)$ gravity theory do have a de Sitter type solution, indicating that for late times the vacuum Universe enters into an exponentially accelerating phase with $q=-1$. From a mathematical point of view the generalized Friedmann equations are given by a set of three highly nonlinear ordinary differential equations, which generally can be solved only numerically. To simplify the numerical analysis we have reformulated the cosmological equations by introducing a set of dimensionless variables. In order to facilitate comparison with observations we have introduced as the independent variable the cosmological redshift $z$.
In our investigations of the Weyl type $f(Q,T)$ gravity we have also analyzed three distinct classes of cosmological models, obtained by choosing some specific simple functional forms for the function $f(Q,T)$. In two of our examples we have considered that $Q$ and $T$ enter in an additive form in the structure of $f(Q,T)$. We have also analyzed a model in which the function $f$ is proportional to the cross term product of $Q$ and $T$, so that  $f\propto QT$. In all three cases we have compared the predictions of the Weyl type $f(Q,T)$ gravity theory with the results of the standard $\Lambda$CDM cosmological model.

 The $f(Q,T)=\alpha Q+\beta T$ model can give a good description of the cosmological data up to redshifts of the order of $z\approx 1-1.5$. Depending on the numerical values of the model parameters  a large variety of cosmological scenarios can be constructed, including cosmological evolutions of the de Sitter type, with the Universe expanding exponentially.  The model $f(Q,T)=\alpha Q\beta T$ also leads to a good description of the standard cosmological model at small redshifts, allowing by an appropriate
choice of the model parameters the construction of a large number of accelerating scenarios, including de Sitter type expansions. The third model with $f(Q,T)=\eta e^{\mu Q}+\nu T$ leads to a complex cosmological dynamics, involving larger deviations from the standard $\Lambda$CDM model. In particular the Universe experiences a very rapid transition from a decelerating phase with a large positive value of $q$ to an accelerating state with $q<0$, and it can reach very quickly a de Sitter type expansion.

 In the Weyl type $f(Q,T)$ gravity theory the nature of the cosmological evolution is strongly dependent on the numerical values of the model parameters, as well as of the functional form of $f$.  For the specific models and the range of cosmological parameters we have considered we have obtained the basic result that the Universe began its recent evolution in a decelerating phase, entering in the large time limit  $z=0$ into an accelerating de Sitter type stage. By slightly varying the model parameters we can obtain a large spectrum of present day values for the deceleration parameter. In general at low redshifts the theoretical predictions of the Hubble parameter in the Weyl type $f(Q,T)$ are similar to those of the standard general relativistic cosmology in the presence of a cosmological constant. However, at higher redshifts significant differences appear with respect to the $\Lambda$CDM model in the behavior of the Hubble function, of the matter energy density, and of the deceleration parameter. However, if investigated for a larger range of functional forms of $f$ and of model parameters the Weyl type $f(Q,T)$ gravity may represent an attractive alternative to the $\Lambda$CDM cosmology, with the late time de Sitter phase induced by the presence of the Weyl geometry, and its interaction with matter.

The Weyl type $f(Q,T)$  gravity theory can be easily generalized to include in the total action, together with ordinary matter, scalar fields. Hence this opens the possibility of another application of the Weyl type $f(Q,T)$ theory, namely, the consideration of inflation in the presence of both Weyl type vector fields, and of scalar fields. Such an approach may lead us to a completely new understanding of the gravitational, geometrical, and cosmological processes that did determine the dynamics  of the very early Universe.  Another major topic that could be investigated in the framework of the Weyl type $f(Q,T)$ gravity is cosmological structure formation, an analysis that could be done with the use of a  background metric. For different choices of the $f(Q,T)$ function the SNIa, BAO, and CMB shift parameter data can be used to obtain constraints for the respective models, and for the evolution of the Weyl vector and of the Lagrange multiplier. Such an approach may also allow the detailed investigation and study of structure formation in the Universe from a different theoretical perspective. Another interesting and important issue is obtaining the Newtonian and the post-Newtonian limits of the Weyl type $f(Q,T)$ gravity theory, an analysis that could allow us to obtain the constraints Solar System level gravity imposes on the theory, and on the properties of the Weyl vector.  Constraints arising from other astrophysical observations can also be obtained by using the Newtonian limit.

In the present paper we have introduced a new model of the symmetric teleparallel theory, in which the nonmetricity is constructed from its initial Weyl form. In this approach the gravitational phenomena can be fully described by the Weyl vector, a Lagrange multiplier, and the metric in a globally flat geometry. By using the variational formulation of the theory we have obtained the basic equations of the model, and we have proven its theoretical consistency. The present results may also motivate and encourage the study the applications of Weyl theory, and of the further extensions of the $f(Q)$ type family of theories. We have also shown that the cosmology of the Weyl type $f(Q,T)$ theory predicts a de Sitter type expansions of the Universe, and it can give a satisfactory description of the cosmological observations usually interpreted in the framework of the standard $\Lambda$CDM model.  Thus the Weyl type $f(Q,T)$ gravity theory may represent a geometric alternative to dark energy, and perhaps even dark matter. In the present study we have proposed some basic theoretical methods for the investigation of the geometric aspects of gravity, and of their astrophysical and cosmological implications.

\section*{Acknowledgments}

We would like to thank the anonymous reviewer for comments and suggestions that helped us to improve our manuscript. T. H. would like to thank the Yat Sen School of the Sun Yat Sen University in Guangzhou, P. R. China, for the kind hospitality offered during the preparation of this work.

\appendix

\section{Derivation of the field equations} \label{EOMderive}

In the present Appendix we will present the detailed calculations used for the derivation of the field equations of the Weyl type $f(Q,T)$ gravity theory.

\subsection{Basic mathematical results}

In this Subsection, we present some basic mathematical results necessary for the derivation of the field equations.
 First of all, we compute $\delta Q/\delta g^{\mu}$ and $\delta T/\delta g^{\mu\nu}$,
 \begin{equation}\label{delta Q}
     \begin{split}
         \frac{\delta Q}{\delta g^{\mu\nu}}&=\frac{\delta}{\delta g^{\mu\nu}}\lp -6w_{\mu}w_{\nu}g^{\mu\nu}\rp=-6w_{\mu}w_{\nu},
     \end{split}
 \end{equation}
 \begin{equation}\label{delta T}
     \frac{\delta T}{\delta g^{\mu\nu}}=\frac{\delta (g^{\alpha\beta}T_{\alpha \beta})}{\delta g^{\mu\nu}}=-T_{\mu\nu}+g_{\mu\nu}L_m,
 \end{equation}
where we have made use of Eq.~(\ref{Q}).

Next we show that $\delta W_{\rho \sigma}/\delta g^{\mu\nu} =0$:
\begin{equation*}
    \frac{\delta}{\delta g^{\mu\nu}}(\nabla_\sigma w_\rho)=\frac{\delta}{\delta g^{\mu\nu}}(\partial_\sigma w_\rho -\Gamma^\tau_{\sigma \rho}w_\tau)=-\frac{\delta \Gamma^\tau_{\sigma \rho}}{\delta g^{\mu\nu}}w_\tau.
\end{equation*}
Similarly, we have
\begin{equation*}
    \frac{\delta }{\delta g^{\mu\nu}}(\nabla_\rho w_\sigma)=-\frac{\delta \Gamma^\tau_{\rho\sigma}}{\delta g^{\mu\nu}}w_\tau.
\end{equation*}

Since the tensor $\delta \Gamma^\tau_{\sigma\rho}$ is symmetric with respect to the two lower indices, we obtain
\begin{equation}\label{delta W}
    \frac{\delta W_{\rho \sigma}}{\delta g^{\mu\nu}}=\frac{\delta}{\delta g^{\mu\nu}}(\nabla_\sigma w_\rho-\nabla_\rho w_\sigma)=0.
\end{equation}

\subsection{Variation with respect to the metric}

 For future convenience we define some terms as follows

\begin{equation*}
    \begin{split}
        I_1& \equiv\frac{\delta}{\delta g^{\mu\nu}}\lb\int d^4x \sqrt{-g}f(Q,T) \rb,\\
        I_2&\equiv\frac{\delta}{\delta g^{\mu\nu}}\int d^4x \sqrt{-g}\lp\lambda \bar{R}\rp,\\
        S_{\mu\nu}&\equiv -\frac{2}{\sqrt{-g}}\frac{\delta}{\delta g^{\mu\nu}}\lb\sqrt{-g}\lp-\frac{1}{4}W_{\alpha\beta}W^{\alpha\beta}-\frac{m^2}{2}w_\alpha w^\alpha\rp\rb,\\
        T_{\mu\nu}&\equiv-\frac{2}{\sqrt{-g}}\frac{\delta(\sqrt{-g}\mathcal{L}_m)}{\delta g^{\mu\nu}}.
    \end{split}
\end{equation*}

Here $T_{\mu\nu}$ and $\tilde{S}_{\mu\nu}$ are the energy-momentum tensors of the matter fields, and the vector field, respectively. A straightforward calculation gives,
\begin{equation}
    \begin{split}
        S_{\mu\nu}&=-\frac{g_{\mu\nu}}{4}W_{\alpha \beta}W^{\alpha\beta}-\frac{m^2}{2}g_{\mu\nu}w^2\\
        &-2\frac{\delta}{\delta g^{\mu\nu}}\lp-\frac{1}{4} W_{\alpha \beta}W^{\alpha \beta}-\frac{m^2}{2}w^2\rp\\
        &=W_{\mu\rho}W_{\nu}^{~\rho}-\frac{g_{\mu\nu}}{4}W_{\alpha \beta}W^{\alpha\beta}+ m^2\lp w_\mu w_\nu-\frac{g_{\mu\nu}}{2}w^2\rp.
    \end{split}
\end{equation}

\subsubsection{Computing $I_1$}
  For the term $I_1$ we find
\bea
        &&I_1=\frac{\delta}{\delta g^{\mu\nu}}\lp \int d^4x \sqrt{-g} f\rp\nonumber\\
        &&=\int d^4x \lp \frac{\delta \sqrt{-g}}{\delta g^{\mu\nu}}f+\sqrt{-g} f_Q \frac{\delta Q}{\delta g^{\mu\nu}}+\sqrt{-g}f_T \frac{\delta T}{\delta g^{\mu\nu}}\rp\nonumber\\
        &&=\sqrt{-g}\lb -\frac{f}{2}g_{\mu\nu}-6f_Q w_\mu w_\nu +f_T(T_{\mu\nu}+\Theta_{\mu\nu})\rb,
\eea
where we have made use of Eqs.~(\ref{delta Q}) and (\ref{delta T}), and of the following identity,
\begin{equation}
    \delta \sqrt{-g}=-\frac{\sqrt{-g}}{2}g_{\mu\nu}\delta g^{\mu\nu}.
\end{equation}
\subsubsection{Computing $I_2$}

For the term $I_2$ we obtain
\begin{equation}
    \begin{split}
        \int d^4x & \delta \lp \sqrt{-g}\lambda \bar{R}\rp=\int d^4x \sqrt{-g}\lp-\frac{1}{2}g_{\mu\nu}\lambda \bar{R}\delta g^{\mu\nu}+\lambda \delta \bar{R}\rp\\
        &=\int d^4x \sqrt{-g}\lambda \lb\delta R+6\delta(\nabla^\rho w_{\rho})-12w_\mu w_\nu \delta g^{\mu\nu} \rb,
    \end{split}
\end{equation}
where we have made use of the fact that $\bar{R}=0$.
 With the use of the identity $\delta R=R_{\mu\nu}\delta g^{\mu\nu}-g_{\mu\nu}\square \delta g^{\mu\nu}-\nabla_{\mu}\nabla_{\nu}\delta g^{\mu\nu}$, and after integrating by parts, we obtain
\begin{equation}
    \begin{split}
        \int d^4x & \sqrt{-g}\lambda \delta R=\\&\int d^4x\sqrt{-g}\lp \lambda R_{\mu\nu}+g_{\mu\nu}\Box \lambda-\nabla_\mu \nabla_\mu\lambda\rp \delta g^{\mu\nu}.
    \end{split}
\end{equation}

Using the identity
\begin{equation}
\delta \Gamma^\alpha_{\beta \gamma}=\frac{1}{2}g^{\alpha \rho}\lp\nabla_\beta\delta g_{\rho\gamma}+\nabla_{\gamma}\delta g_{\rho \beta}-\nabla_\rho \delta g_{\beta \gamma} \rp,
\end{equation}
and the fact that $g^{\alpha \beta}\nabla_\nu g_{\beta\gamma}+g_{\beta \gamma}\nabla_\mu g^{\alpha \beta}=\nabla_\mu {\delta^\beta _\gamma}=0$, we obtain
\begin{equation}
    \begin{split}
        &\int d^4x \sqrt{-g}\lambda \delta \lp\nabla_\alpha w^\alpha \rp\\
        &=\int d^4x\sqrt{-g}\lambda \lb\nabla_\alpha w_\beta \delta g^{\alpha\beta}+g^{\alpha\beta}\delta\lp\nabla_\alpha w_\beta \rp\rb\\
        &=\int d^4x \sqrt{-g}\lambda \lb\nabla_\alpha w_\beta \delta g^{\alpha\beta}-g^{\alpha\beta}w_{\gamma}\delta \Gamma^{\gamma}_{\alpha\beta}\rb\\
        &=\int d^4x \sqrt{-g}\lambda \Big[ \nabla_\alpha w_\beta \delta g^{\alpha\beta}\\&-\frac{1}{2}w^\rho g^{\beta\alpha}\lp\nabla_\alpha \delta g_{\beta \rho}+\nabla_\beta \delta g_{\alpha\rho}-\nabla_\rho \delta g_{\alpha\beta}\rp\Big]\\
        &=\int d^4x \sqrt{-g}\lambda \Big[ \nabla_\alpha w_\beta \delta g^{\alpha\beta}\\&+\frac{1}{2}w^\rho\lp g_{\beta\rho}\nabla_\alpha \delta g^{\beta \alpha}+g_{\alpha\rho}\nabla_\beta \delta g^{\alpha\beta}-g_{\alpha\beta}\nabla_\rho \delta g^{\alpha\beta}\rp\Big]\\
        &=\int d^4x \sqrt{-g}\lb\lambda \nabla_\alpha w_\beta -\nabla_\alpha (\lambda w_\beta)+\frac{1}{2}g_{\alpha\beta}\nabla_\rho(\lambda w^\rho)\rb\\
        &=\int d^4x \sqrt{-g}\lb-w_{(\beta} \nabla_{\alpha)} \lambda+\frac{1}{2}g_{\alpha\beta}\nabla_\rho(\lambda w^\rho)\rb.
    \end{split}
\end{equation}
Hence:
\begin{equation}
\begin{split}
    I_2&=\sqrt{-g}\Big[\lambda R_{\mu\nu}+g_{\mu\nu}\Box \lambda-\nabla_\mu \nabla_\nu \lambda-6\lambda w_\mu w_\nu\\&-6w_{(\mu}\nabla_{\nu )}\lambda
    +3g_{\mu\nu}\nabla_\rho (\lambda w^\rho)\Big ].
    \end{split}
\end{equation}
According to the least action principle,
\begin{equation}
    \frac{\delta S}{\delta g^{\mu\nu}}=\kappa^2 I_1+I_2-\f12\sqrt{-g}(T_{\mu\nu}+S_{\mu\nu})=0,
\end{equation}
and therefore we arrive at the field equation
\begin{align}\label{EOM21}
&\f12\left( T_{\mu\nu}+S_{\mu\nu}\right)-\kappa^2f_T \left(T_{\mu\nu}+\Theta_{\mu\nu}\right)=-\frac{\kappa^2}{2}g_{\mu\nu}f\nonumber\\
&-6\kappa^2f_Q w_\mu w_\nu+\lambda \left(R_{\mu\nu}-6w_\mu w_\nu+3g_{\mu\nu}\nabla_\rho w^\rho\right)\nonumber\\
& +3g_{\mu\nu} w^\rho \nabla_\rho \lambda   -6w_{(\mu}\nabla_{\nu)}\lambda
+g_{\mu\nu}\Box\lambda -\nabla_\mu \nabla_\nu \lambda,
\end{align}

\subsection{Variation with respect to $w_\mu$}

Since the matter energy momentum tensor is independent of the vector field, we have
\begin{equation}
    \frac{\delta f}{\delta w_\mu}=f_Q \frac{\delta Q}{\delta w_\mu}=-12 f_Q w^\mu,
\end{equation}
where we have made use of Eq.~(\ref{Q}). Now it is easy to check the following results,
\begin{equation}
    \begin{split}
        &\frac{\delta\lp W_{\alpha\beta}W^{\alpha\beta}\rp}{\delta w_\mu}=4\lp\Box w^\mu-\nabla_\nu \nabla^\mu w^\nu\rp=4\nabla_\nu W^{\mu\nu},\\
        &\frac{\delta w^2}{\delta w_\mu}=2w^\mu,\\
        &\lambda \frac{\delta \bar{R}}{\delta w_\mu}=6\lambda \frac{\delta (\nabla_\alpha w^\alpha)}{\delta w^\mu}-6\lambda \frac{\delta w^2}{\delta w^\mu}=-6\nabla_\mu \lambda-12 \lambda w_\mu.
    \end{split}
\end{equation}
Thus the field equation of the Weyl type vector field is
\begin{equation}\label{EOM11}
\nabla^\nu W_{\mu \nu }-(m^2+12\kappa^2 f_Q+12\lambda)w_\mu=6\nabla_\mu \lambda.
\end{equation}
\section{Alternative form of the field equations}\label{appB}

In this Appendix, we present the derivation of Eqs.~(\ref{simpEOM2}).
Firstly, we introduce some notations as
\begin{equation*}
    \begin{split}
        &D^{(0)}_{\mu\nu}=R_{\mu\nu}-6w_\mu w_\nu+3g_{\mu\nu}\nabla_\rho w^\rho,\\
        &D^{(1)}_{\mu\nu}=3g_{\mu\nu}w^\rho \nabla_\rho-6w_{(\mu}\nabla_{\nu)},\\
        &D^{(2)}_{\mu\nu}=g_{\mu\nu}\Box-\nabla_\mu\nabla_\nu,\\
        &D_\mu=\nabla_\mu+2w_\mu.
    \end{split}
\end{equation*}
Using the above definitions, Eq.~(\ref{EOM2}) can be written as
\bea
    \f12\left(T_{\mu\nu}+S_{\mu\nu}\right)&&-\kappa^2f_T\left(T_{\mu\nu}+\Theta_{\mu\nu}\right)=-\frac{\kappa^2}{2}g_{\mu\nu}f\nonumber\\-6\kappa^2f_Q w_\mu w_\nu
    &&+(D^{(0)}_{\mu\nu}+D^{(1)}_{\mu\nu}+D^{(2)}_{\mu\nu})\lambda.
    \eea

From Eq.~(\ref{EOM1}), we see that $D_\mu \lambda$ is independent of $\lambda$, so that
\begin{equation}
    D_\mu \lambda=\frac{1}{6}\nabla^\nu W_{\nu\mu}-\lp\frac{m^2}{6}+2\kappa^2f_Q\rp w_\mu.
\end{equation}

Therefore we can replace ordinary derivatives with $D_\mu$, and try to eliminate all the derivatives of lambda. Using the following identities,
\begin{align}
        &\nabla_\mu \lambda=D_\mu\lambda-2w_\mu \lambda,\\
        &\nabla_\nu \nabla_\mu\lambda=\nabla_\nu D_\mu \lambda-2\nabla_\nu (w_\mu \lambda),\\
        &\Box \lambda=\nabla^\mu D_\mu\lambda-2\lambda \nabla_\mu w^\mu-2w_\mu \nabla^\mu \lambda,
\end{align}
and starting from the highest order derivative, we have
\bea
    D^{(2)}_{\mu\nu}\lambda&=&g_{\mu\nu} \nabla^\alpha D_\alpha \lambda-\nabla_\mu D_\nu \lambda-2g_{\mu\nu}w_\rho \nabla^\rho \lambda\nonumber\\
    &&+\left(2\nabla_\mu (\lambda w_\nu)-2\lambda g_{\mu\nu}\nabla_\rho w^\rho\right).
    \eea

Now, we obtain
\bea
    &&\left(D^{(0)}_{\mu\nu}+D^{(1)}_{\mu\nu}+D^{(2)}_{\mu\nu}\right)\lambda =  g_{\mu\nu}\nabla^\rho D_\rho \lambda-\nabla_\nu D_\mu\lambda\nonumber\\
    &&
    +g_{\mu\nu}w_\rho\nabla^\rho \lambda-w_\mu\nabla_\nu \lambda-3w_\nu\nabla_\mu \lambda\nonumber\\
    &&
    +\lambda \lp R_{\mu\nu}-6w_\mu w_\nu+g_{\mu\nu}\nabla_\rho w^\rho+2\nabla_\mu w_\nu\rp.
\eea
Working on the first order derivatives in a similar manner, we obtain
\begin{align}
        &(D^{(0)}_{\mu\nu}+D^{(1)}_{\mu\nu}+D^{(2)}_{\mu\nu})\lambda = g_{\mu\nu}\nabla^\rho D_\rho \lambda-\nabla_\nu D_\mu\lambda\nonumber\\
        &
    +g_{\mu\nu}w_\rho D^\rho \lambda-w_\mu D_\nu \lambda-3w_\nu D_\mu \lambda\nonumber\\
    &
    +\lambda \lp R_{\mu\nu}+2w_\mu w_\nu-2g_{\mu\nu}w^2+g_{\mu\nu}\nabla_\rho w^\rho+2\nabla_\nu w_\mu\rp.
\end{align}
Hence the field equations become
\begin{align}
  \f12&\left(T_{\mu\nu}+S_{\mu\nu}\right)-\kappa^2f_T\left(T_{\mu\nu}+\Theta_{\mu\nu}\right)=-\frac{\kappa^2}{2}g_{\mu\nu}f\nonumber\\&-6\kappa^2f_Q w_\mu w_\nu+  g_{\mu\nu}\nabla^\rho D_\rho \lambda-\nabla_\nu D_\mu\lambda\nonumber\\
  &
  + g_{\mu\nu}w_\rho D^\rho \lambda-w_\mu D_\nu \lambda-3w_\nu D_\mu \lambda\nonumber\\
  &
  +\lambda \lp R_{\mu\nu}+2w_\mu w_\nu-2g_{\mu\nu}w^2+g_{\mu\nu}\nabla_\rho w^\rho+2\nabla_\nu w_\mu\rp.
\end{align}

\end{document}